\documentclass{aa}
\usepackage{graphicx,natbib}
\usepackage[latin1]{inputenc}
\usepackage{color}
\bibpunct{(}{)}{;}{a}{}{,}
\usepackage{txfonts}
\usepackage{changebar,ulem}

\def\mcfost{{\sf MCFOST}}

\begin{document}

\title{Planet gaps in the dust layer of 3D protoplanetary disks:}
\subtitle{II. Observability with ALMA}
\titlerunning{Planet gaps in the dust layer of 3D protoplanetary disks. II}
\author{J.-F. Gonzalez \inst{1}
       \and 
       C. Pinte \inst{2}
       \and
       S. T. Maddison \inst{3}
       \and
       F. M\'enard \inst{2}
       \and
       L. Fouchet \inst{4}}
\institute{Universit\'e de Lyon, Lyon, F-69003, France ; Universit\'e Lyon~1, Observatoire de Lyon, 9 avenue Charles Andr\'e, Saint-Genis Laval, F-69230, France ; CNRS, UMR 5574, Centre de Recherche Astrophysique
de Lyon ; \'Ecole Normale Sup\'erieure de Lyon, Lyon, F-69007, France\\
        \email{Jean-Francois.Gonzalez@ens-lyon.fr}
        \and
         UJF-Grenoble 1 / CNRS-INSU, Institut de Plan\'etologie et d'Astrophysique de Grenoble, UMR 5274, Grenoble, F-38041, France\\
        \email{christophe.pinte@obs.ujf-grenoble.fr,francois.menard@obs.ujf-grenoble.fr}	
        \and
        Centre for Astrophysics and Supercomputing, Swinburne Institute of Technology, PO Box 218, Hawthorn, VIC 3122, Australia\\
        \email{smaddison@swin.edu.au}
        \and
        Physikalisches Institute, Universit\"at Bern, CH-3012 Bern, Switzerland\\
        \email{laure.fouchet@space.unibe.ch}
}

\date{Received 11 January 2012 / Accepted 17 August 2012}

\abstract
{The Atacama Large Millimeter/submillimeter Array (ALMA) will have the necessary resolution to observe a planetary gap created by a Jupiter-mass planet in a protoplanetary disk. Because it will observe at submillimeter and millimeter wavelengths, grains in the size range 10\,$\mu$m to 1\,cm are relevant for the thermal emission. For the standard parameters of a T~Tauri disk, most grains of this size range are weakly coupled to the gas (leading to vertical settling and radial migration) and the common approximation of well-mixed gas and dust does not hold.}
{We provide predictions for ALMA observations of planet gaps that account for the specific spatial distribution of dust that results from consistent gas+dust dynamics.}
{In a previous work, we ran full 3D, two-fluid Smoothed Particle Hydrodynamics (SPH) simulations of a planet embedded in a gas+dust T~Tauri disk for different planet masses and grain sizes. In this work, the resulting dust distributions are passed to the Monte Carlo radiative transfer code \mcfost\ to construct synthetic images in the ALMA wavebands. We then use the ALMA simulator to produce images that include thermal and phase noise for a range of angular resolutions, wavelengths, and integration times, as well as for different inclinations, declinations and distances. We also produce images which assume that gas and dust are well mixed with a gas-to-dust ratio of 100 to compare with previous ALMA predictions, all made under this hypothesis.}
{Our findings clearly demonstrate the importance of correctly incorporating the dust dynamics. We show that the gap carved by a 1\,$M_\mathrm{J}$ planet orbiting at 40\,AU is visible with a much higher contrast than the well-mixed assumption would predict. In the case of a 5\,$M_\mathrm{J}$ planet, we clearly see a deficit in dust emission in the inner disk, and point out the risk of interpreting the resulting image as that of a transition disk with an inner hole if observed in unfavorable conditions. Planet signatures are fainter in more distant disks but declination or inclination to the line-of-sight have little effect on ALMA's ability to resolve the gaps.}
{ALMA has the potential to see signposts of planets in disks of nearby star-forming regions. We present optimized observing parameters to detect them in the case of 1 and 5\,$M_\mathrm{J}$ planets on 40\,AU orbits.}

\maketitle

\section{Introduction}
\label{sec:Introduction}

The study of exoplanets has steadily gained momentum in the past two decades, in particular since the confirmed identification of the first exoplanet around the solar-like star 51~Peg \citep{MayorQueloz95}. More than 700 planets located in $\sim$500 planetary systems are now confirmed\footnote{\texttt{http://exoplanet.eu}}. A direct consequence of these surveys is that more reliable estimations of the statistical properties of planetary systems are becoming available (e.g., distributions of orbital distances and planet masses, see \citealp{Schneider2011}). 

Two scenarios currently compete for giant gaseous planet formation: core-accretion \citep[e.g.][]{AMBW05} and disk gravitational instability \citep[e.g.][and references therein]{Boss2011}, each with its own merits and limitations. In the core-accretion scenario, giant planets form in two steps. A rocky core is first formed by progressive accumulation of planetesimals until a critical mass ($\sim$10\,$M_\oplus$) is reached. Then rapid capture of nebular gas can occur, in a runaway fashion, until the planet reaches its final mass. In the disk instability scenario, a gravitationally unstable disk fragments into self-gravitating clumps which then directly form gas giant planets. The relevance of each mechanism to explain the current observations of exoplanets, especially at large orbital radii, depends largely on the exact properties of the underlying protoplanetary disk \citep[see e.g.][]{Mordasini2012}. 

Although circumstellar disks of gas and dust were imaged directly roughly at the same time as the exoplanets \citep{Dutrey1994, Burrows1996} it is only recently that the link between them and the exoplanets was confirmed directly by observations when planets and disks were imaged simultaneously around $\beta$~Pictoris \citep{Lagrange2010} and HR~8799 \citep{Marois2008,Marois2010}. Both systems are however rather evolved and gas giant formation has likely ceased in these now gas-poor debris disks. Interestingly, the case of HR~8799 with several planets orbiting between 14 and 68\,AU suggests that the formation of giant planets on wide orbits is a strong possibility. Indeed, \citet{Quanz2012} recently found that a larger fraction of stars than previously assumed may harbor gas giant planets at larger orbital separations and \citet{Lambrechts2012} showed that such planets can be formed by the core accretion mechanism well within the disk lifetime.

ALMA, when completed, will have the capacity to observe these wide planetary systems at a younger age and explore a parameter space that is complementary to that probed by radial velocity and transit techniques. In this paper we show that signs of the presence of these planets on wide orbits, and specifically the gaps they carve in the disk, will be detectable with ALMA. The process of gap formation depends on the planet mass, the disk surface density and the size of dust grains \citep{PM04,Fouchet07} and measuring their width and depth can provide constraints on the underlying disk structure \citep{CMM06}. Previous predictions indicate that ALMA should be able to detect such gaps \citep{W02,W05}. Most of these predictions were made assuming the very best performance ALMA will provide, i.e., longest baselines, shortest wavelengths, no or low phase noise.  The dust responsible for the emission was also assumed to be well-mixed with the gas, irrespective of the dust particle size. Here we will use the results of numerical simulations \citep[hereafter Paper~I]{Fouchet10} where the dust distributions are calculated for each grain size using a two-phase SPH approach to produce ALMA synthetic images. We will assess the different planet signatures due to different dust distributions as a function of the grain size. We will explore the ability of ALMA to detect these gaps, but also to measure the disk properties both inside and outside the planet's orbit for various combinations of observing time, angular resolution and wavelength in different star forming regions in order to estimate the optimum survey strategy.

\defcitealias{Fouchet10}{Paper~I}

\section{Methods}
\label{sec:Methods}

In this section, we present the three numerical tools which we use to produce synthetic ALMA images of disks with planetary gaps. We first introduce our hydrodynamic code, from which we obtain the three-dimensional dust and gas distributions for a protoplanetary disk with an embedded planet. This dust distribution is fed into a radiative transfer code to produce raw intensity maps. Finally, to make predictions of what ALMA will see for a given wavelength, angular resolution, and integration time for disks at a specific distance, declination and inclination, we use the ALMA simulator from the CASA\footnote{\texttt{http://casa.nrao.edu}} (Common Astronomy Software Applications) package.

\subsection{Hydrodynamic simulations}
\label{sec:Hydro}

In a previous work \citepalias{Fouchet10}, we ran simulations of a disk of mass 0.02\,$M_\odot$ extending from 4 to 160\,AU around a solar-mass star, typical of Classical T~Tauri Stars. These were run with our two-phase 3D SPH code \citep{BF05} comprising gas and dust, in which the dust is treated as a pressureless fluid. The initial dust-to-gas ratio is set to 0.01. Both phases interact via aerodynamic drag in the Epstein regime. Grain sizes are fixed (grain growth and fragmentation are neglected) and self-gravity is not included in this study. We use a vertically isothermal temperature profile, $T=T_0 r^{-1}$, with an initial disk aspect ratio of $H/r=0.05$. The initial density profile is flat, $\Sigma(r)=\Sigma_0$.

\begin{figure}
\centering
\resizebox{\hsize}{!}{
\includegraphics{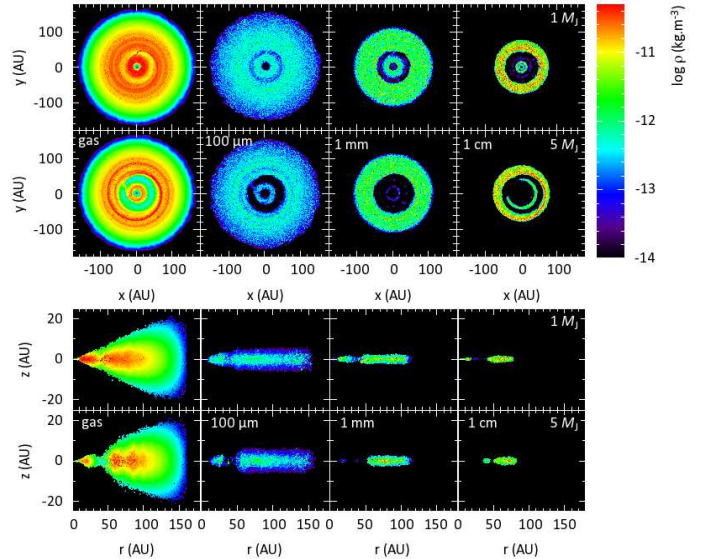}
}
\caption{Results of the hydrodynamic simulations: volume density maps in midplane (top) and meridian plane (bottom) cuts of the disks. The leftmost column shows the gas density and the three rightmost columns show the dust density for 100\,$\mu$m, 1\,mm and 1\,cm grains, from left to right. The rows show simulations with $M_\mathrm{p}=1$ and $5\,M_\mathrm{J}$, from top to bottom. The colorbar applies to all plots.}
\label{Fig:hydro}
\end{figure}

We embed a planet on a circular orbit at 40\,AU in a gas-only disk, which we evolve for 8~planetary orbits. We then inject the dust phase by overlaying dust particles on the gas and evolve the system for a total of 104 planetary orbits, following the influence of the planet on the vertical settling and radial migration of solids. Simulations were ran for different grain sizes: 100\,$\mu$m, 1\,mm, 1\,cm (smaller grains being well coupled to the gas) and planet masses: 0.1, 0.5, 1 and 5\,$M_\mathrm{J}$. For more details, see \citetalias{Fouchet10}. In this work, we use the simulation results for planet masses of 1 and 5\,$M_\mathrm{J}$. Fig.\,\ref{Fig:hydro} shows the results of the hydrodynamic simulations for both planet masses and all three grain sizes.

In \citetalias{Fouchet10}, we showed that the resulting gap is deeper and wider for larger grains, as well as for more massive planets. The 1\,$M_\mathrm{J}$ planet only carves a shallow gap in the gas while it opens a deep gap in the dust phase. The more massive 5\,$M_\mathrm{J}$ planet carves a deep gap in both the gas and dust phases and also traps the 1\,cm grains in corotation inside the gap (visible as the horseshoe-shaped feature in Fig.\,\ref{Fig:hydro}), whereas they pile up along the spiral wave in the outer disk, causing an asymmetric structure. Different radial migration efficiencies for different grain sizes cause the varying outer disk radius and inner disk dust density (in this paper, we call inner and outer disk the disk regions interior and exterior to the planet gap, resp.). An extensive discussion of the resulting dynamics can be found in \citetalias{Fouchet10}.

\subsection{Coupling of the hydro simulations and radiative transfer code}
\label{sec:Coupling}

Synthetic images are computed using \mcfost, a 3D continuum radiative transfer code based on the Monte Carlo method \citep{Pinte2006,Pinte09}. The code propagates photon packets whose transport is governed by successive scattering, absorption and re-emission events determined by the local dust properties. The temperature structure is calculated using the immediate re-emission algorithm of \cite{Bjorkman01} combined with a continuous deposition of energy to estimate the mean intensity \citep{Lucy99}. We assume dust grains to be homogeneous spheres (Mie theory) and adopt the porous dust grain optical properties from \citet[][model A]{Mathis89}, with an average grain density of 0.5\,g\,cm$^{-3}$.

To produce synthetic thermal emission maps, we need a prescription of the three-dimensional dust distribution at every point in the disk and thus we need to transform the SPH particle description to a grid description. This transformation is directly performed on the \mcfost\ spatial grid. The dust mass in any given cell of the radiative transfer code is obtained by summing the mass of the neighbouring SPH particles, multiplied by the corresponding SPH kernels (see App.~\ref{App:N_SPH} for a discussion of the sampling). We use the distributions obtained from our SPH simulations for the 100\,$\mu$m, 1\,mm and 1\,cm grains. Grains smaller than 100\,$\mu$m, whose spatial distribution was not simulated in \citetalias{Fouchet10}, are strongly coupled to the gas. We therefore assume that all grains of 10$\,\mu$m and smaller are well mixed with the gas and follow its spatial distribution.
 
Following this procedure, we obtain the spatial density for 4~grain sizes (10$\,\mu$m and below, 100$\,\mu$m, 1\,mm and 1\,cm). We then compute the spatial densities of grains of all other sizes $a$ between $a_\mathrm{min}=0.03\,\mu$m and $a_\mathrm{max}=1$\,cm by performing a linear interpolation in the $(\log a,\log\rho)$ plane. The relative amount of each grain size is obtained by normalizing the densities so that the grain size distribution integrated over the whole disk follows $\mathrm{d}n(a) \propto a^{-3.5}\,\mathrm{d}a$, and the total dust mass is equal to 0.01 of the gas mass. We call this setup the ``dynamic'' case. This procedure has been successfully applied to study dust vertical settling in the disk of GG~Tau \citep{Pinte2007}.

In order to assess the impact of gas and dust decoupling on the simulated ALMA images, we also produce grain size distributions under the assumption that grains of all sizes are well mixed with the gas. In this case, we simply use the resulting SPH gas spatial distribution as a proxy to any grain distribution and call this setup the ``well-mixed'' case.
 
The inner radius of 4\,AU only has a limited impact on the disk temperature structure since the disk is optically thick to stellar radiation in the radial direction. In the regions probed by ALMA ($> 15$\,AU) that we discuss in this work, the heating is due to reprocessing of the stellar radiation absorbed in the surface layers. The temperature structure at large scales is therefore not affected by the exact value of the inner radius (see App.~\ref{App:Tstruct}). As a consequence, only the central beam of the synthetic ALMA map depends on our inner radius of 4\,AU.

\subsection{Producing synthetic ALMA images}
\label{sec:CASA}

Images of the disk as it would be observed by ALMA are computed from the \mcfost\ maps using the CASA ALMA simulator (release 3.2, build 15111). The simulator first computes complex visibilities at each of the ($u$,$v$) points sampled by the considered ALMA configuration and then adds thermal and phase noise before reconstructing an image using the CLEAN algorithm. Thermal noise is computed by constructing an atmospheric model from the site characteristics with a given value of Precipitable Water Vapor (PWV) and a ground temperature of $270$\,K. The atmospheric transparency above Chajnantor has been monitored over the period 1973--2003 \citep{ALMA512}: the quartiles of the distribution of the 225\,GHz sky optical depth over this period are $\tau_{225}=0.0383$, 0.0625 and 0.1174. The CASA simulator uses a linear relation between $\tau_{225}$ and the PWV. The corresponding PWVs for the 3 quartiles are 0.58, 1.08 and 2.22\,mm. The best 10\% of sky conditions corresponds to $\tau_{225}\simeq0.025$ and a PWV of 0.30\,mm. These data are summarized in Table\,\ref{Tab:SkyConditions}. In the following, we mostly focus on the median sky quality and use 1.08\,mm of PWV unless otherwise stated.

\begin{table}
\caption{Sky conditions at Chajnantor from \citet{ALMA512}: percentiles of the distribution of the 225\,GHz sky optical depth $\tau_{225}$ with its corresponding value and the equivalent amount of precipitable water vapor (PWV).}
\label{Tab:SkyConditions}
\begin{center}
\begin{tabular}{lll}\hline\hline
Percentile & $\tau_{225}$ & PWV (mm) \\ \hline
10\% & 0.025  & 0.30 \\
25\% & 0.0383 & 0.58 \\
50\% & 0.0625 & 1.08 \\
75\% & 0.1174 & 2.22 \\ \hline
\end{tabular}
\end{center}
\end{table}

Because the troposphere is refractive, atmospheric turbulence introduces phase delays which are not uniform between antennas. This will primarily affect the most extended configurations and highest frequencies. Two nearby antennas will roughly see the same part of the atmosphere and differential phase delays will be small, whereas well-separated antennas see different parts of the atmosphere with uncorrelated deformations, resulting in larger phase delays. The larger opacity of water vapor at higher frequencies also results in larger phase delays. We use the CASA tool \textit{settrop} to build a 2-dimensional phase screen according to a fractional Brownian motion, allowing us to create realistic correlations in time and space between antennas. In practice, this creates a fluctuation of precipitable water vapor screen (we assume a relative amplitude of 15\%) which is used to calculate the corresponding phase delay as a function of frequency. The phase screen is blown across the array with a wind speed of 7\,m\,s$^{-1}$. ALMA will use Water Vapor Radiometers (WVRs) to correct for the atmospheric phase delay in real time during the observation.
Their efficiency is not yet fully known, and in the following we present simulations with and without phase noise. We do not include the correction by the WVRs. The quality of the ALMA data will lie between these two extreme cases, depending on the quality of this correction.

For this work, we have chosen to present reconstructed images from simulated ALMA observations rather than visibility plots. This is common in the literature \citep[e.g.][]{Cossins2010,Regaly2012} and therefore easier to compare our results with previous predictions of planet gap observations \citep{W02,W05}. Images are also a convenient way to reveal the instrument signature when the input disk structure is known, especially in the favorable case of nearly pole-on disks. Finally, the imaging capabilities offered by a large number of antennas has been a driver for the building of ALMA. However, observers having obtained ALMA data, and without any a priori knowledge of the source, will mostly rely on visibilities to understand its structure \citep[similarly to, e.g.,][]{Hughes2007}.

\subsection{Parameter space}
\label{sec:ParameterSpace}

For each of the synthetic maps computed with \mcfost, we produce a grid of ALMA images for a range of observational parameters. We choose a reference disk with an inclination $i=18.2^\circ$ (the first in the set of 10 equally spaced values in cosine used in Sect.\,\ref{sec:Inclination}) at a declination of $-23^\circ$ and a distance of 140\,pc, and vary the integration time (from 10\,min to 8\,h) and angular resolution (from $0.05''$ to $0.5''$) for four different wavelengths: 350\,$\mu$m, 850\,$\mu$m, 1.3\,mm and 2.7\,mm. We then determine the optimal observing mode for all wavelengths (where optimal means best compromise between $S/N$, reasonable integration time and angular resolution which recover the features of the disk), and then use this observing mode to compare images from the well-mixed and dynamic cases at different wavelengths for both planet masses. We follow by varying the disk inclination and finally, we investigate the effects of source declination and distance to determine the observability of planetary gaps in different star-forming regions. Table\,\ref{Tab:ImageParam} lists the full set of parameters we explored to produce all the ALMA simulated images presented in this paper.

\begin{figure}
\centering
\resizebox{\hsize}{!}{
\includegraphics{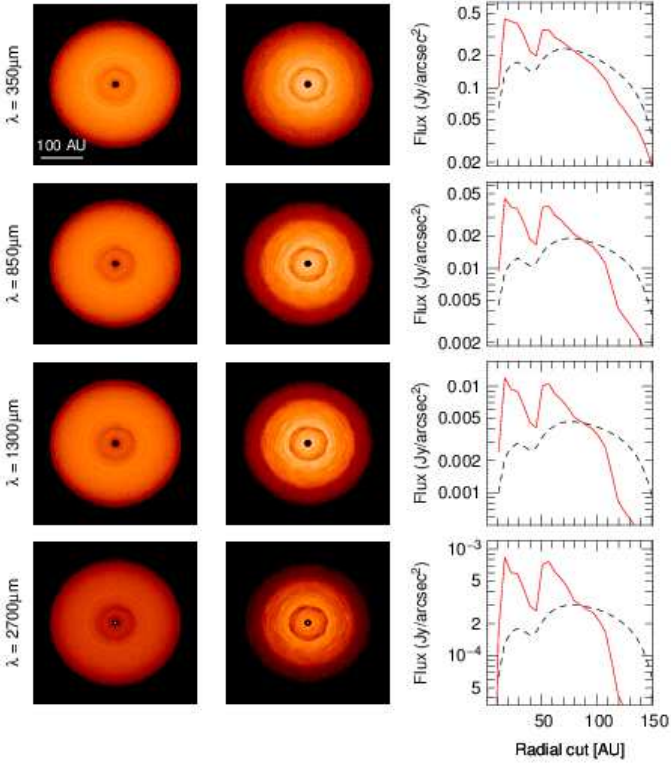}
}
\caption{Raw synthetic images for the 1\,$M_\mathrm{J}$ planet. Left: well-mixed case, center: dynamic case, right: radial brightness profiles of the disks in the well-mixed case (dashed black line) and the dynamic case (solid red line). From top to bottom, $\lambda=350\,\mu$m, 850\,$\mu$m, 1.3\,mm and 2.7\,mm.}
\label{Fig:1Mj_raw}
\end{figure}

\begin{figure}
\centering
\resizebox{\hsize}{!}{
\includegraphics{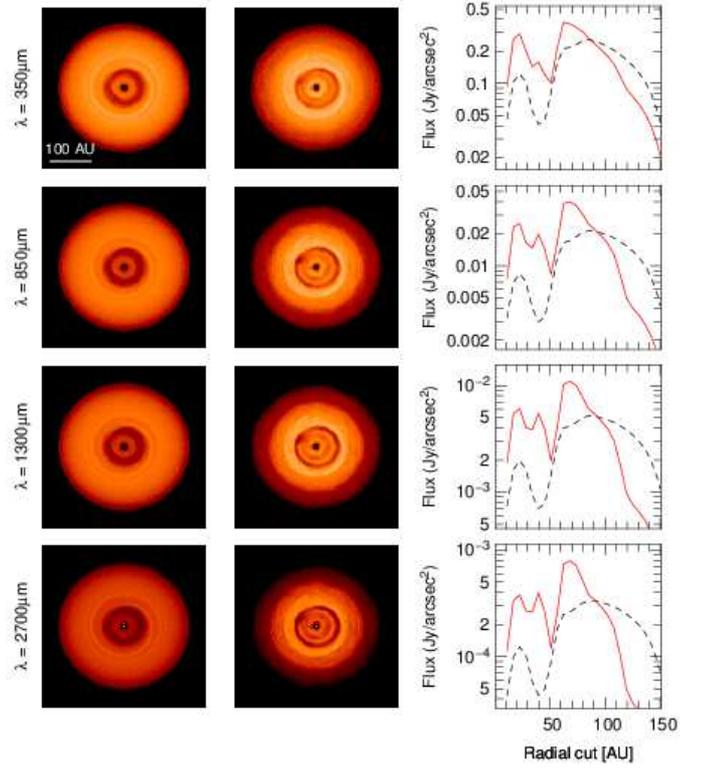}
}
\caption{Same as Fig.\,\ref{Fig:1Mj_raw} for the 5\,$M_\mathrm{J}$ planet.}
\label{Fig:5Mj_raw}
\end{figure}

\section{Results and discussion}
\label{sec:Results}

\subsection{Raw synthetic images}
\label{sec:SynthIm}

In Figs.\,\ref{Fig:1Mj_raw} and \ref{Fig:5Mj_raw}, we present the raw synthetic images produced by \mcfost\ for both the well-mixed and dynamic cases for four different wavelengths and both planet masses. An obvious difference between the well-mixed and dynamic results is that the extent of the dust disk varies with wavelength in the dynamic case but remains fixed for the well-mixed case. The results of the hydrodynamic simulations shown in Fig.\,\ref{Fig:hydro} demonstrate that the radial extent of the dust layer varies with grain size, due to the varying migration efficiency with grain size. In the synthetic images, all the grain sizes contribute to each wavelength, but the grains with sizes closest to the observing wavelength will contribute the most. We therefore naturally see the smaller disk radius for the longer wavelength in the dynamic case.

In the case of the 1\,$M_\mathrm{J}$ planet, the gap is much more visible in the dynamic case than in the well-mixed case (Fig.\,\ref{Fig:1Mj_raw}). This is mainly due to the increase in brightness at the gap edges, which results in a stronger contrast in the dynamic than in the well-mixed case. The hydrodynamic simulations show that this results from the pile-up of dust grains in the pressure maxima at the gap edges. The brightness contrast increases with wavelength in the dynamic case because at longer wavelengths the larger grains, which are more effectively cleared out of the gap, are the dominant contributors. In the well-mixed case, grains of all sizes follow the gas and the resulting contrast is therefore the same for all wavelengths. The same contrast effect is seen for the 5\,$M_\mathrm{J}$ planet between the gap and the outer disk (Fig.\,\ref{Fig:5Mj_raw}). For the more massive planet, another obvious difference between the well-mixed and dynamic models is that features characteristic of one particular grain size have different relative contributions as the wavelength varies in the dynamic case. The inner disk, which only contains 100\,$\mu$m grains, naturally appears brighter at shorter wavelengths. (The larger grains migrate more efficiently and hence have almost all been accreted onto the star, see Fig.\,\ref{Fig:hydro}.) Similarly, the particles in corotation, which comprise only 1\,cm grains, appear brighter at longer wavelengths.

\begin{figure*}
\resizebox{\hsize}{!}{
\includegraphics[angle=-90]{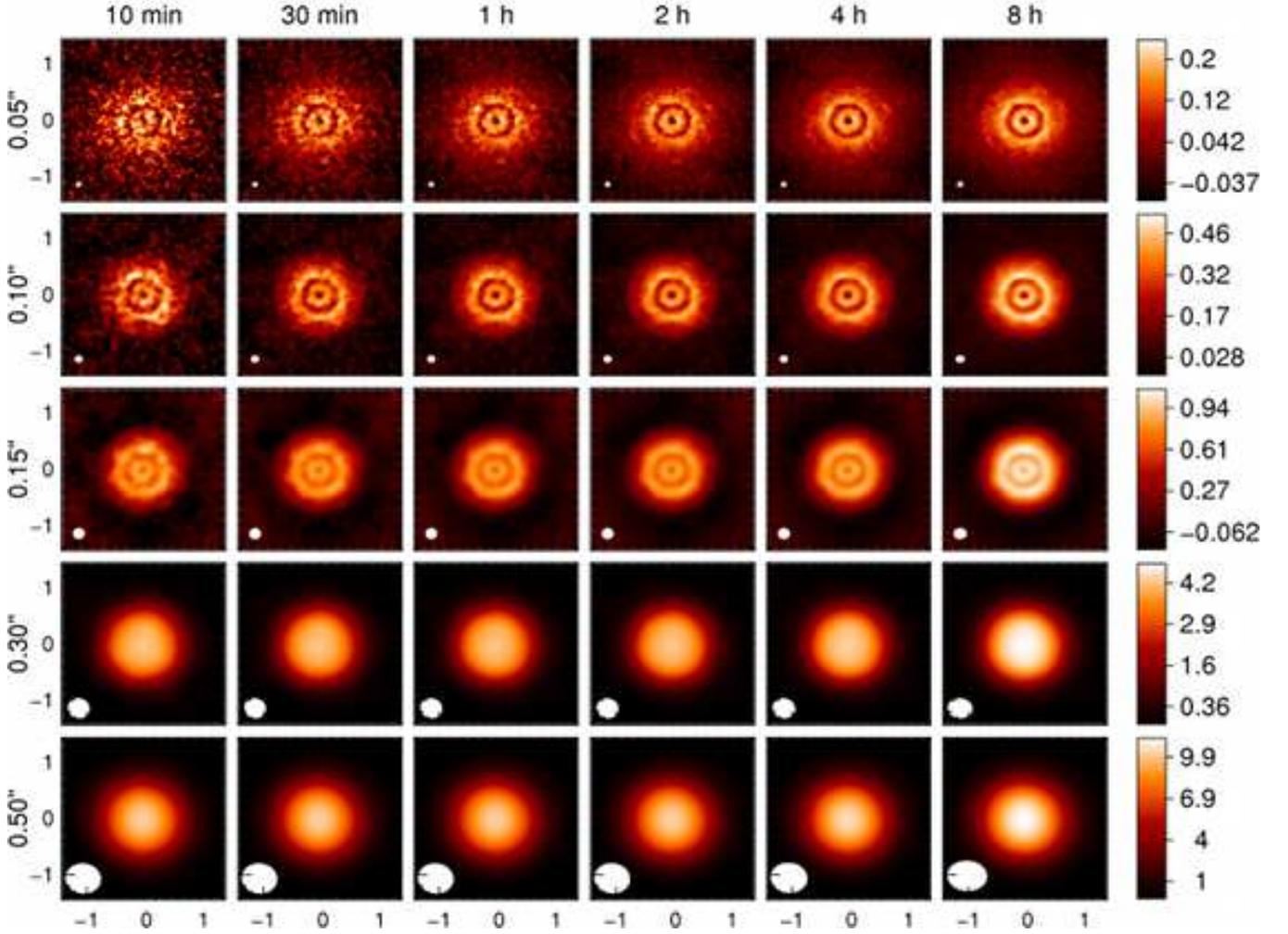}
}
\caption{Simulated observations for the 1\,$M_\mathrm{J}$ planet in the dynamic case at $\lambda=850\ \mu$m. From left to right: integration time of 10\,min, 30\,min, 1\,h, 2\,h, 4\,h and 8\,h. From top to bottom: angular resolution of $0.05''$, $0.10''$, $0.15''$, $0.30''$ and $0.50''$ (the ALMA configuration is adapted to match the angular resolution for each wavelength). The scale on each image is in arcseconds, with the beam size represented at its bottom left corner, and the colorbar gives the flux in mJy/beam. (Note that the flux scale changes in each row, due to the different beam size.)}
\label{Fig:IntTimeResol}
\end{figure*}

Although both planets produce a spiral wave in the gas phase, which is mirrored in the raw images for the well-mixed case, it is weaker and barely visible for the 1\,$M_\mathrm{J}$ planet. In the dust phase, grains accumulate in the pressure maxima along the spiral wave \citepalias[see][]{Fouchet10}, producing correspondingly stronger features in the images for the dynamic case. No significant enhancement of the weak spiral is discernible for the 1\,$M_\mathrm{J}$ planet, but the 5\,$M_\mathrm{J}$ planet produces a stronger spiral structure in the images for the dynamic case.

While we can see a range of different features in the dynamic vs. well-mixed cases, one must remember that these are raw synthetic images. In order to predict what a real telescope will see, we now turn to results from the ALMA simulator. 

\begin{table*}
\caption{Parameters used to produce all ALMA simulated images presented in this paper's figures.}
\label{Tab:ImageParam}
\begin{center}
\begin{tabular}{lllllllll}\hline\hline
Fig. & $M_\mathrm{p}$ ($M_\mathrm{J}$) & Case\tablefootmark{a} & $\lambda$ ($\mu$m) & Integration time (h) & Angular resolution ($''$) & $i$ ($^\circ$) & PWV (mm) & Phase noise \\ \hline
\ref{Fig:IntTimeResol} & 1 & D & 850 & 1/6, 1/2, 1, 2, 4, 8 & 0.05, 0.10, 0.15, 0.30, 0.50 & 18.2 & 1.08 & No \\
\ref{Fig:Contrast} & 1, 5 & D & 350--2700 & 1/6--8 & 0.10 & 18.2 & 1.08 & No \\
\ref{Fig:1Mj_CASA} & 1 & WM,D & 350, 850, 1300 & 1 & 0.10 & 18.2 & 1.08 & No \\
\ref{Fig:5Mj_CASA} & 5 & WM,D & 350, 850, 1300 & 1 & 0.10 & 18.2 & 1.08 & No \\
\ref{Fig:1Mj_PhaseNoise} & 1 & D & 350, 850, 1300 & 1 & 0.10 & 18.2 & 1.08, 0.58, 0.30 & Yes \\
\ref{Fig:5Mj_PhaseNoise} & 5 & D & 350, 850, 1300 & 1 & 0.10 & 18.2 & 1.08, 0.58, 0.30 & Yes \\
\ref{Fig:Contrast_PhaseNoise} & 1, 5 & D & 350--2700 & 1/6--8 & 0.10 & 18.2 & 1.08 & Yes \\
\ref{Fig:Inclination} & 1 & D & 850 & 1 & 0.10 & 18.2--87.1 & 1.08 & No \\
\ref{Fig:SFR} & 1 & D & 850 & 1 & adapted\tablefootmark{b} & 18.2 & 1.08 & No \\
\ref{Fig:ALMAlimits} & 1, 5 & D & 350, 850, 1300 & 8 & 0.05 & 18.2 & 1.08 & No \\
\ref{Fig:ALMAlimitsPWV} & 1, 5 & D & 350 & 8 & 0.05 & 18.2 & 0.30, 0.58 & No \\
\ref{Fig:Contrast_PWV0.3} & 1, 5 & D & 350--2700 & 1/6--8 & 0.05 & 18.2 & 0.30 & No \\
\hline
\end{tabular}
\tablefoot{\tablefoottext{a}{WM: well-mixed, D: dynamic}
\tablefoottext{b}{see Sect.\,\ref{sec:OtherSFRs}}}
\end{center}
\end{table*}

\subsection{ALMA synthetic images}
\label{sec:ALMApredictions}

We start by investigating the effect of the integration time and angular resolution on ALMA's ability to resolve features in disks with planet gaps. In Fig.\,\ref{Fig:IntTimeResol}, we show as an example the ALMA simulated images for the 1\,$M_\mathrm{J}$ planet at $\lambda=850\ \mu$m, for exposure times ranging from 10\,min to 8\,h and for angular resolutions from $0.05''$ to $0.50''$. Increasing the integration time increases the sensitivity and therefore the signal-to-noise ratio. In particular, the background becomes less noisy as the integration time increases. However, the gap as well as other features such as brightness variation in the disk are already well detectable after 1\,h and integrating longer does not produce much additional detail. An integration time of 1\,h thus seems to be the best compromise between being able to detect the essential properties of the disk and using a reasonable amount of resources, especially in the context of an observatory that is well oversubscribed.

\begin{figure*}
\resizebox{.5\hsize}{!}{
\includegraphics{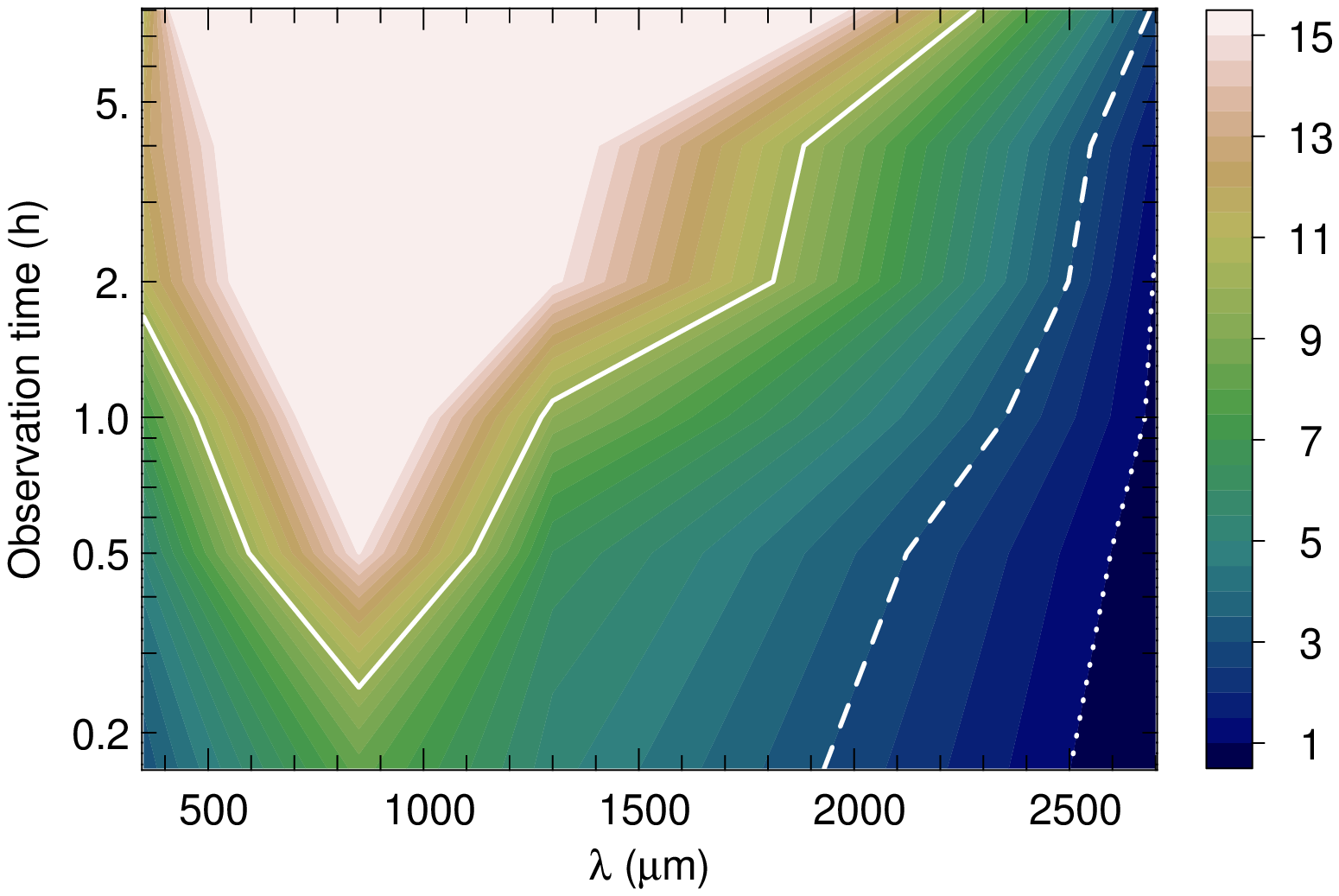}
}
\resizebox{.5\hsize}{!}{
\includegraphics{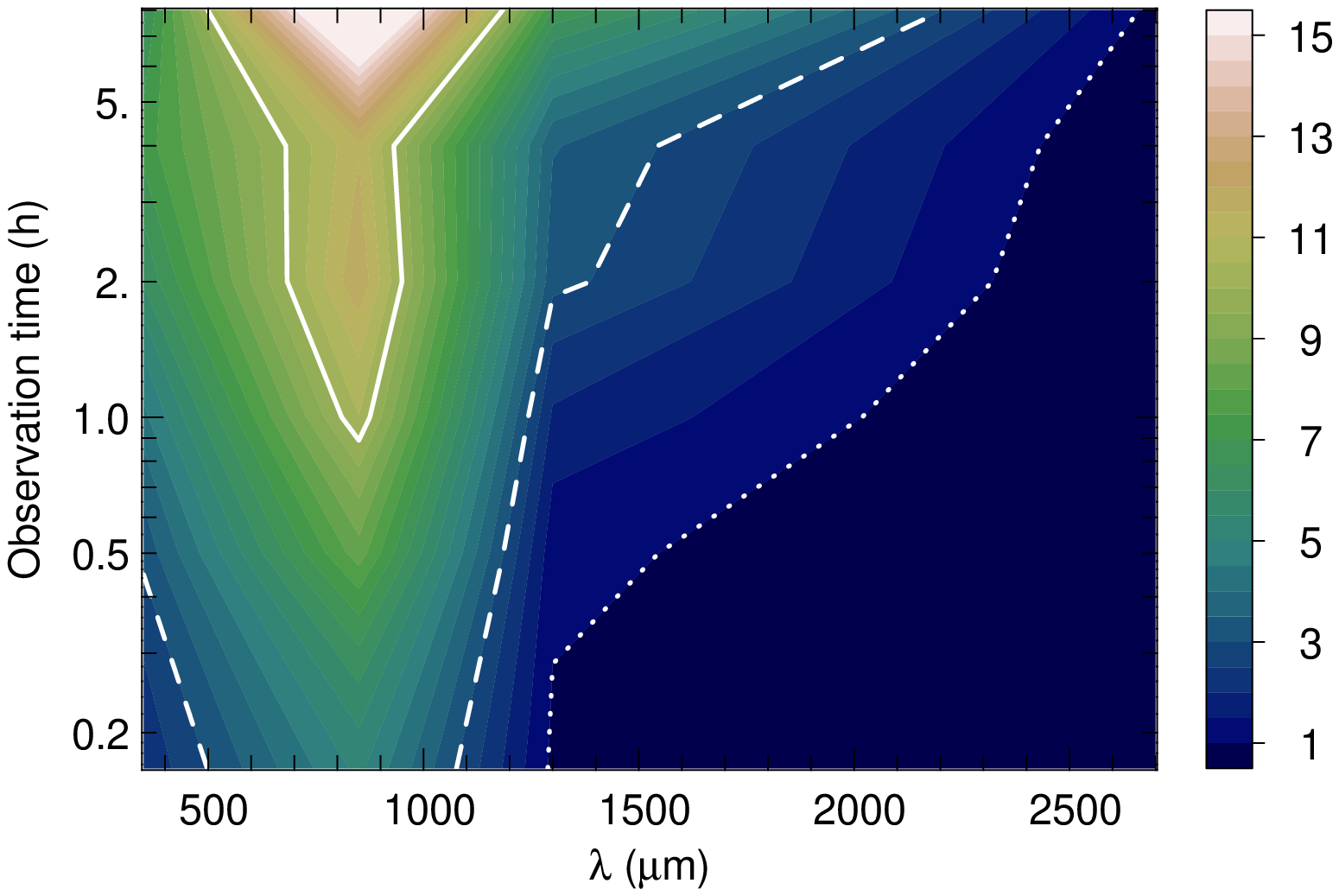}
}
\resizebox{.5\hsize}{!}{
\includegraphics{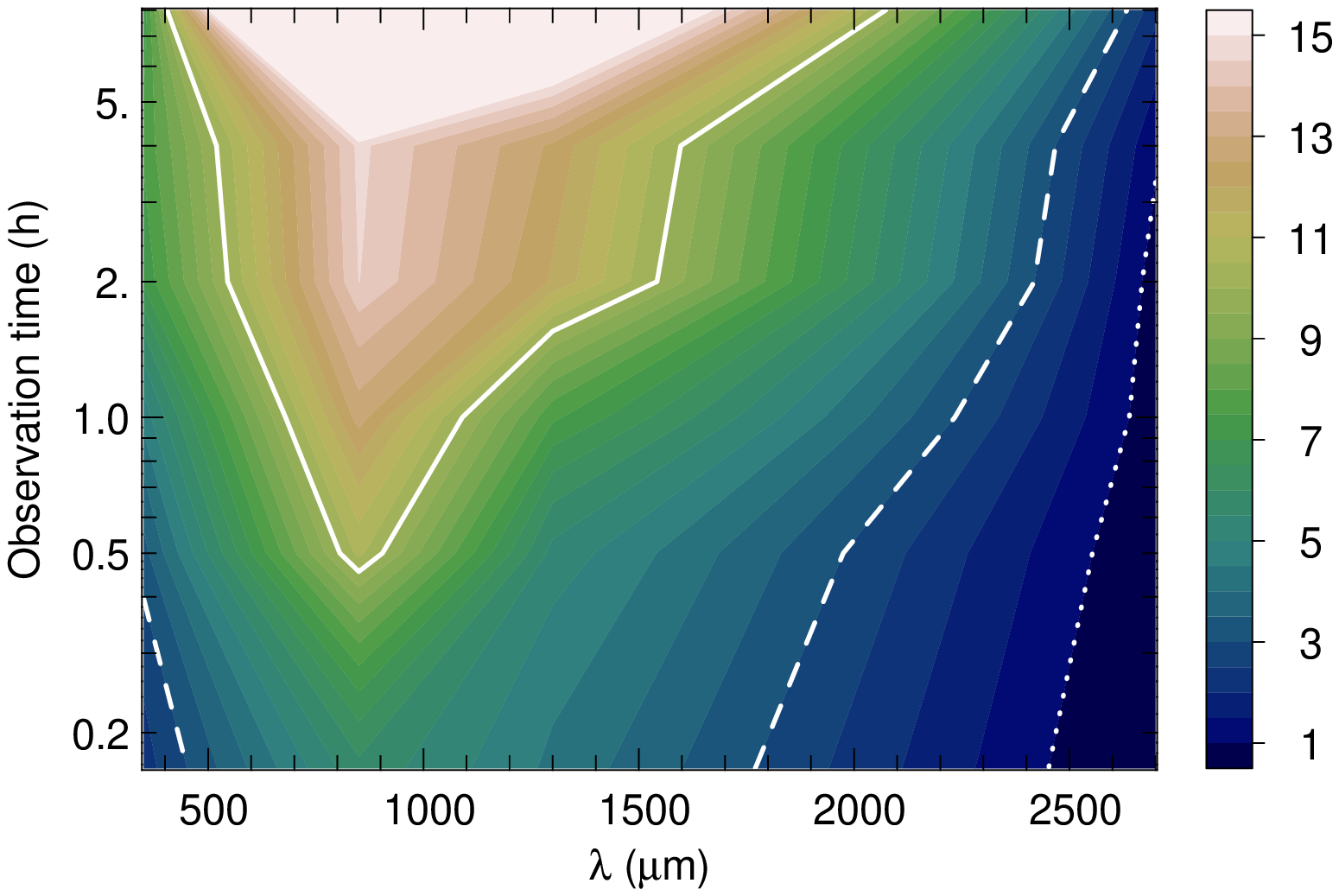}
}
\resizebox{.5\hsize}{!}{
\includegraphics{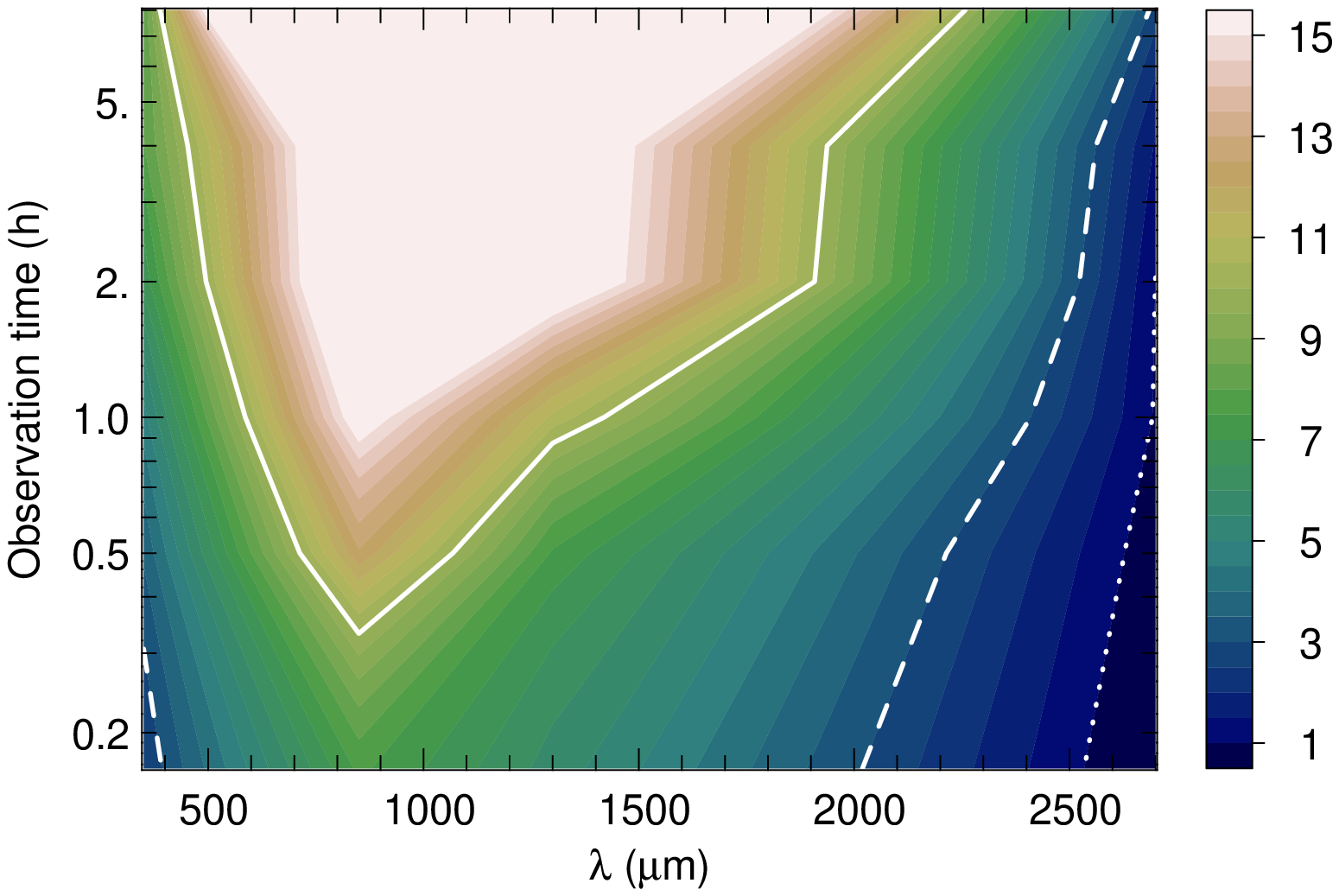}
}
\caption{Maps of the signal-to-noise ratio in the inner (top) and outer (bottom) disks in the wavelength-integration time plane for the dynamic case, with an angular resolution of $0.10''$. The signal-to-noise ratio is computed as the median of the flux in a ring (between $0.07"$ and $0.21"$ for the inner disk and between $0.35"$ and $0.70"$ for the outer disk) divided by the rms of the background. Left: $M_\mathrm{p}=1\,M_\mathrm{J}$, right: $M_\mathrm{p}=5\,M_\mathrm{J}$. Thick dotted, dashed and solid lines show contours for $S/N=1$, 3 and 10.}
\label{Fig:Contrast}
\end{figure*}

Figure\,\ref{Fig:IntTimeResol} shows that an angular resolution of $0.15''$ or smaller is necessary to detect the gap at 850~$\mu$m. However, higher angular resolution requires more extended array configurations, with correspondingly decreasing beam sizes, therefore sampling smaller areas of the source. The extended array would detect less flux, resulting in decreased contrast in the image. We find that a resolution of $0.10''$ gives the best disk-to-gap contrast and adopt this value for the rest of this paper. We verify that the optimal combination of an integration time of 1\,h and angular resolution of $0.10''$ is adequate for all wavelengths studied here, except for 2.7\,mm --- see Figs\,\ref{Fig:IntTimeResol350}-\ref{Fig:IntTimeResol2700}. For the longest ALMA waveband, the minimum angular resolution required to resolve the gap corresponds to a very extended configuration, and our optimal observing parameters result in a very weak signal, making it extremely difficult to detect the disk and its features (see App.\,\ref{App:Figs} showing figures similar to Fig.\,\ref{Fig:IntTimeResol} for the other wavelengths).

Figure\,\ref{Fig:Contrast} shows maps of the signal-to-noise ratio per resolution element in the inner and outer disk (i.e.\ the disk regions interior and exterior to the gap, resp.) for both planet masses in the wavelength vs.\ integration time plane for the dynamic case with $0.10''$ angular resolution. These maps, showing very low values of $S/N$ for 2.7\,mm, confirm our statement of the previous paragraph. Only very long integration times would allow the S/N to approach or slightly exceed unity at this wavelength, and we therefore no longer discuss it in the rest of this paper. For shorter wavelengths, Fig.\,\ref{Fig:Contrast} shows that a 1\,h integration time is indeed enough to reach a $S/N$ larger than 7--8, and even well above 10 at 850\,$\mu$m (except in the inner disk for $M_\mathrm{p}=5\ M_\mathrm{J}$, which will be discussed in Sect.\,\ref{sec:Detect}), ensuring an adequate detection of the planetary gaps.

\subsection{``Well-mixed'' vs. ``dynamic''}
\label{sec:WMvsD}

\begin{figure}
\resizebox{\hsize}{!}{
\includegraphics{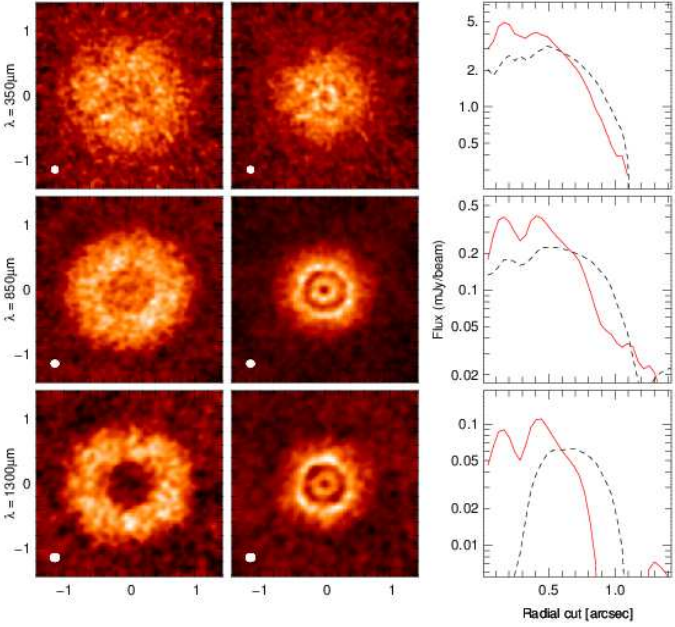}
}
\caption{Simulated observation for the 1\,$M_\mathrm{J}$ planet with an integration time of 1\,h and an angular resolution of $0.10''$. Left: well-mixed case, center: dynamic case, right: radial brightness profiles of the disks in the well-mixed case (dashed black line) and the dynamic case (solid red line). From top to bottom, $\lambda=350\,\mu$m, 850\,$\mu$m and 1.3\,mm (the 2.7\,mm image is not shown because it contains only noise.) The scale on the images and radial profiles is in arcseconds.}
\label{Fig:1Mj_CASA}
\end{figure}

\begin{figure}
\resizebox{\hsize}{!}{
\includegraphics{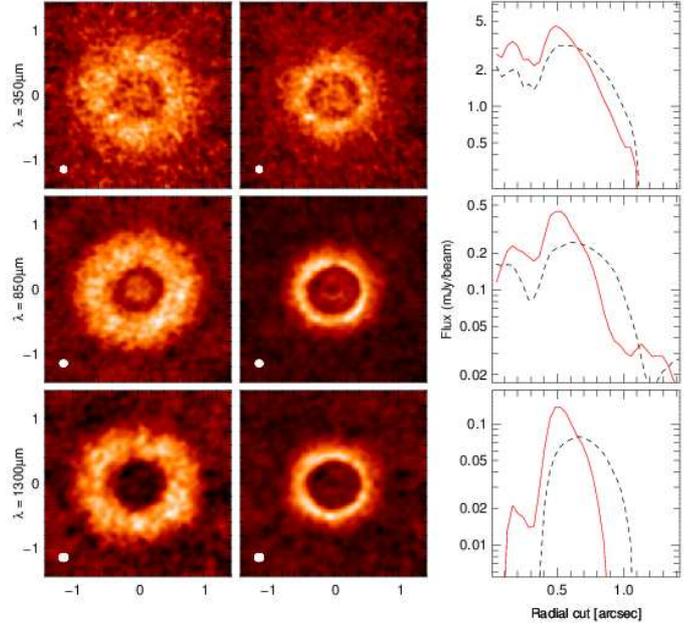}
}
\caption{Same as Fig\,\ref{Fig:1Mj_CASA} for the 5\,$M_\mathrm{J}$ planet.}
\label{Fig:5Mj_CASA}
\end{figure}

We next compare simulated images with our optimal set of observing parameters (1\,h integration time, $0.10''$ angular resolution) of the 1\,$M_\mathrm{J}$ (Fig.\,\ref{Fig:1Mj_CASA}) and 5\,$M_\mathrm{J}$ (Fig.\,\ref{Fig:5Mj_CASA}) planet gaps for the well-mixed and dynamic cases. Quantitative information is obtained from the radial brightness profiles of these images, shown in the right columns of Figs.\,\ref{Fig:1Mj_CASA} and \ref{Fig:5Mj_CASA}. As already seen in the raw synthetic images, the disks are more compact in the dynamic case, and all the more so for longer wavelengths (down to a radius of $\sim$0.8$''$ at 1.3~mm), than in the well-mixed case (with a constant radius of $\sim$1.1$''$). However, a number of features of the raw images are lost here.

In the well-mixed case, the inner disk is too faint to be visible for both planet masses at 1.3\,mm, showing instead a disk with a large inner hole, which is wider for the more massive planet (with radii of $\sim$0.2$''$ and $\sim$0.4$''$ for $M_\mathrm{p}=1$ and 5\,$M_\mathrm{J}$ resp.). At 850\,$\mu$m, the $1\,M_\mathrm{J}$ planet gap is hard to detect due to the shallow ($\sim$15\%) dip and mild slope of its brightness profile. Its location can however be inferred from to the $\sim$20\% brightness contrast between the fainter inner disk and the outer disk. On the other hand, the gap is well defined and $\sim$1.9 times fainter than the inner disk for the $5\,M_\mathrm{J}$ planet. The situation is less favorable at 350\,$\mu$m where the $5\,M_\mathrm{J}$ planet gap is deep enough to be detected ($\sim$1.5 times fainter than the inner disk) but its edges are less sharp, whereas it is hard to identify a gap in the shallow brightness variations for the $1\,M_\mathrm{J}$ planet. No additional details can be seen to provide constraints on the underlying disk structure.

In the dynamic case, where the dust features are sharper, the planet gaps are very well defined. The 1\,$M_\mathrm{J}$ planet gap is clearly detected at all wavelengths, with sharp transitions to both the inner and outer disk (only at 350\,$\mu$m is the transition to the outer disk smoother). The disk-to-gap contrast can be estimated from the right column of Fig.\,\ref{Fig:1Mj_CASA} to be $\sim$1.4 at 350\,$\mu$m, $\sim$1.6 at 850\,$\mu$m and $\sim$2 at 1.3\,mm. For the 5\,$M_\mathrm{J}$ planet, however, the fainter inner disk is clearly visible only at 350\,$\mu$m, whereas it is very faint at 850\,$\mu$m (as already seen in the upper right panel of Fig.\,\ref{Fig:Contrast}) and barely detectable at 1.3\,mm. This, combined with the strong contrast between the outer disk and the gap ($\sim$2.1, $\sim$2.5 and $\sim$10 at 350\,$\mu$m, 850\,$\mu$m and 1.3\,mm) constitutes a possible source of confusion with disks having cleared a large inner hole (see below). Finally, the corotating 1\,cm sized grain population and the spiral wave are no longer visible in the ALMA simulated images, depriving us of a potential handle on the planet mass or angular position.

These notable differences between images computed in both cases constitute one of the main results of our study: the much higher contrast of the planet gaps in the more realistic dynamic case compared to the well-mixed approximation clearly shows that they will be much easier to detect than anticipated from gas-only simulations, such as those of \citet{W02}. This was seen by \citet{PM04} in the case of 2D, face-on disks. As we showed in \citetalias{Fouchet10}, for 3D disks the much reduced scale height of the settled dust phase compared to the gas further enhances the sharpness of the dust features. As a result, they are easier to detect in the simulated images we present here. In studying 3D disks it is also possible to investigate the effect of disk inclination, see Sect.~\ref{sec:Inclination}. In the rest of this paper, we will no longer consider the well-mixed approximation and only discuss images obtained from the dynamic case.

\subsection{Gap detectability}
\label{sec:Detect}

\begin{figure}
\centering
\resizebox{\hsize}{!}{
\includegraphics{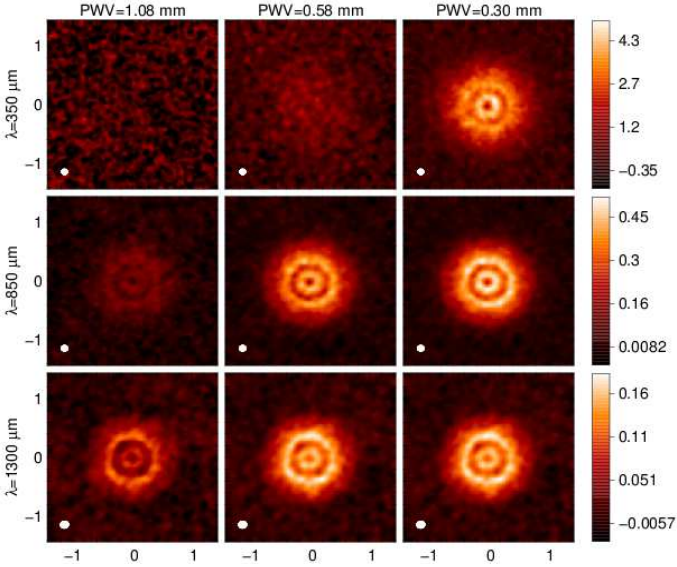}
}
\caption{Simulated observation for the 1\,$M_\mathrm{J}$ planet with an integration time of 1\,h and an angular resolution of $0.10''$ in the dynamic case with phase noise included. Left: PWV\,=\,1.08\,mm, center: PWV\,=\,0.58\,mm, right: PWV\,=\,0.30\,mm. From top to bottom, $\lambda=350\,\mu$m, 850\,$\mu$m, and 1.3\,mm. The scale on each image is in arcseconds, with the beam size represented at its bottom left corner, and the colorbar gives the flux in mJy/beam.}
\label{Fig:1Mj_PhaseNoise}
\end{figure}

We now assess the detectability of the planet gaps in simulated images using our optimal parameters. A single 1-hour ALMA image at a well-chosen wavelength (Figs.\,\ref{Fig:1Mj_CASA} and \ref{Fig:5Mj_CASA} show 850\,$\mu$m to be a good choice) can be sufficient to detect a planet gap and infer the presence of an unseen planet. This is a definite advantage for future large surveys of star-forming regions. However, characterizing the planet which opens the gap will likely require multi-wavelength follow-up observations. First, weighing the planet can not rely solely on the measure of the observed gap width, which does not depend strongly on the planet mass (see the right columns of Figs.\,\ref{Fig:1Mj_CASA} and \ref{Fig:5Mj_CASA}), partly because it (as well as the gap-to-disk flux ratio) is affected by the beam size. One can use in addition the different wavelength dependence of other features produced by planets of different masses, such as the inner disk brightness. In our example with 1 and $5\,M_\mathrm{J}$ planets, the inner disk brightness decreases more rapidly towards longer wavelengths for the more massive planet. Furthermore, in the case where the inner disk is too faint to be unambiguously detected, it is important to be able to differentiate it from a disk with a large inner hole. In recent years, a growing number of disks with large dust-free cavities around the central star have been observed \cite[see e.g.][]{Andrews2011}. The clearing of the inner regions of these so-called transition disks is attributed either to dynamical interactions with a companion star \citep{AL1994} or a massive planet \citep{Varniere2006} or, more efficiently, multi-planet systems \citep{Dodson2011}, or to photoevaporative winds \citep{Alexander2006,Gorti2009}. Our simulated ALMA observations for the 5\,$M_\mathrm{J}$ planet show that the inner disk, while difficult to detect at 850\,$\mu$m and even more so at 1.3\,mm, is easily seen at 350\,$\mu$m (Fig.\,\ref{Fig:5Mj_CASA}). Being able to image the disk at short wavelengths therefore seems essential to disentangle disks with planetary gaps from transition disks in the dust continuum. Even though the presence of warm dust in the inner disk can be hinted at by an IR excess in the Spectral Energy Distribution, imaging the inner disk is essential to constrain the amount of dust it contains. In the case where the inner disk is present but too faint to be unambiguously detected in reconstructed images, visibilities may show emission in the regions close to the star, providing an alternate way of discriminating disks with planetary gaps from transition disks. A different method for doing so has recently been proposed by \citet{Cleeves2011}, who claim that the heated inner wall of transition disks would produce unique chemical features in the gas phase, potentially observable by ALMA.

It is worth noting that gaps in protoplanetary disks can be created without the presence of a planet. Before clearing an inner hole, X-ray photoevaporation opens a gap at a few AU from solar-type stars \citep{Owen2011} but such a feature is short-lived ($\sim$5\% of the disk lifetime, i.e. a few hundred thousand years) before the whole inner disk is drained on the star. In contrast, the planet gaps we consider here take less than 20\,000\,yr to open \citepalias[see][]{Fouchet10} and survive until the disk is cleared. \citet{Pinilla2012} considered the effect of a bumpy surface density profile and showed that the accumulation of dust at local maxima would be seen with ALMA as a succession of rings and gaps. They speculate that such a density profile may be caused by the magneto-rotational instability, however this remains to be tested.

\subsection{Effects of phase noise}
\label{sec:phasenoise}

\begin{figure}
\centering
\resizebox{\hsize}{!}{
\includegraphics{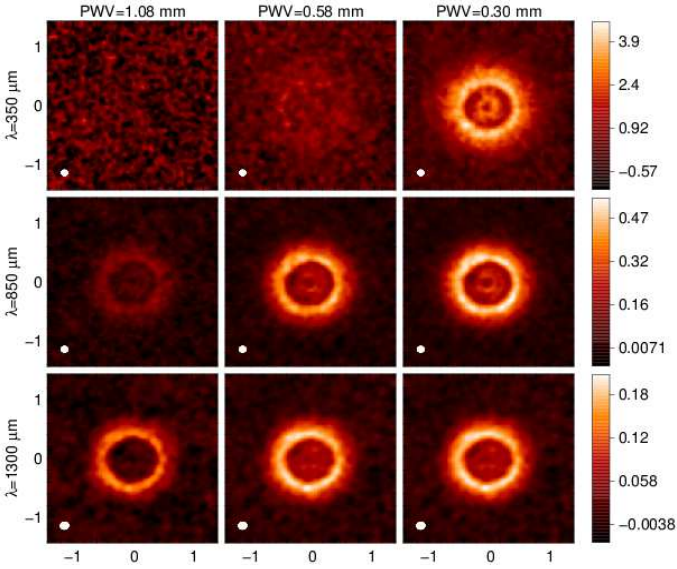}
}
\caption{Same as Fig.\,\ref{Fig:1Mj_PhaseNoise} for the 5\,$M_\mathrm{J}$ planet.}
\label{Fig:5Mj_PhaseNoise}
\end{figure}

\begin{figure*}
\resizebox{.5\hsize}{!}{
\includegraphics{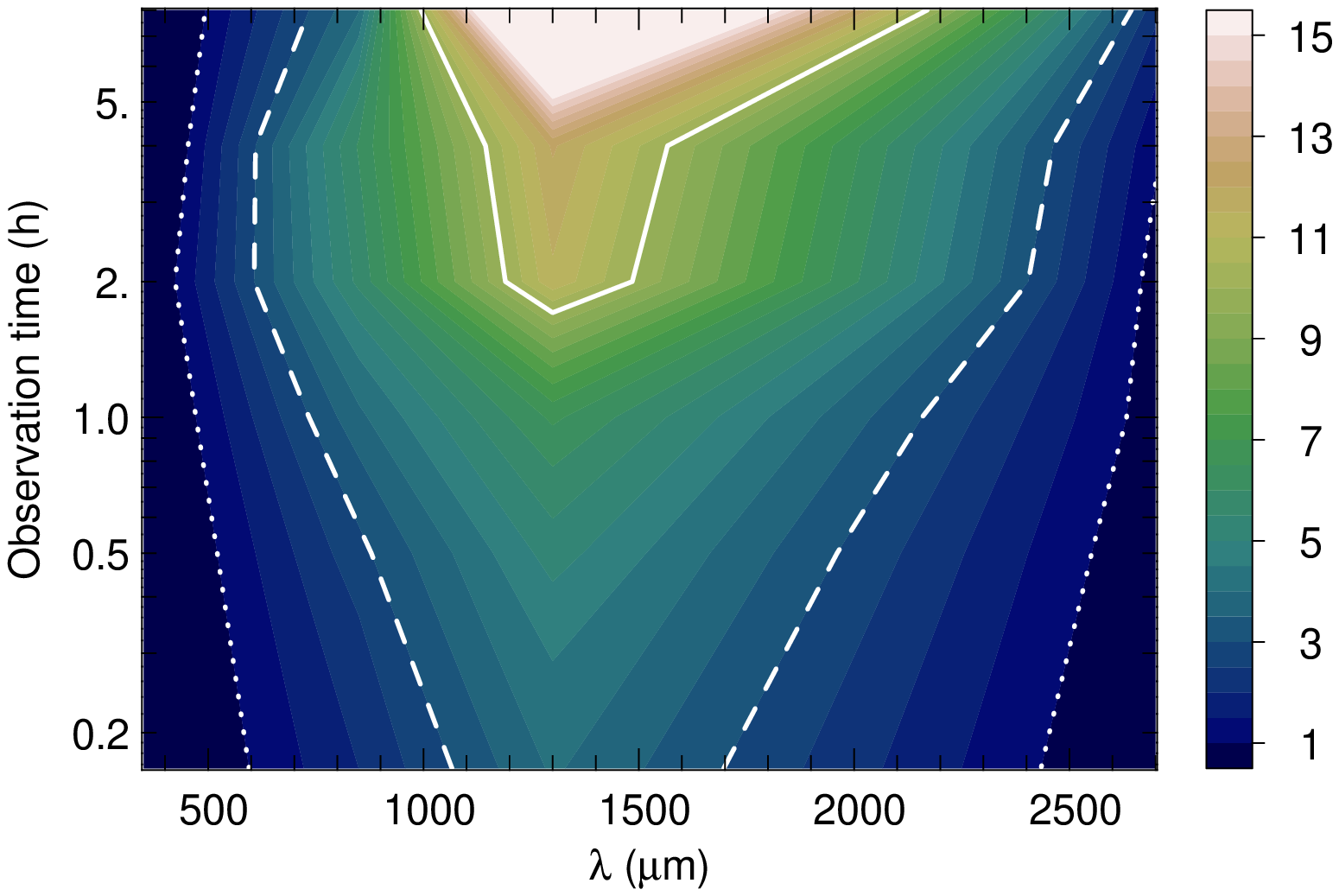}
}
\resizebox{.5\hsize}{!}{
\includegraphics{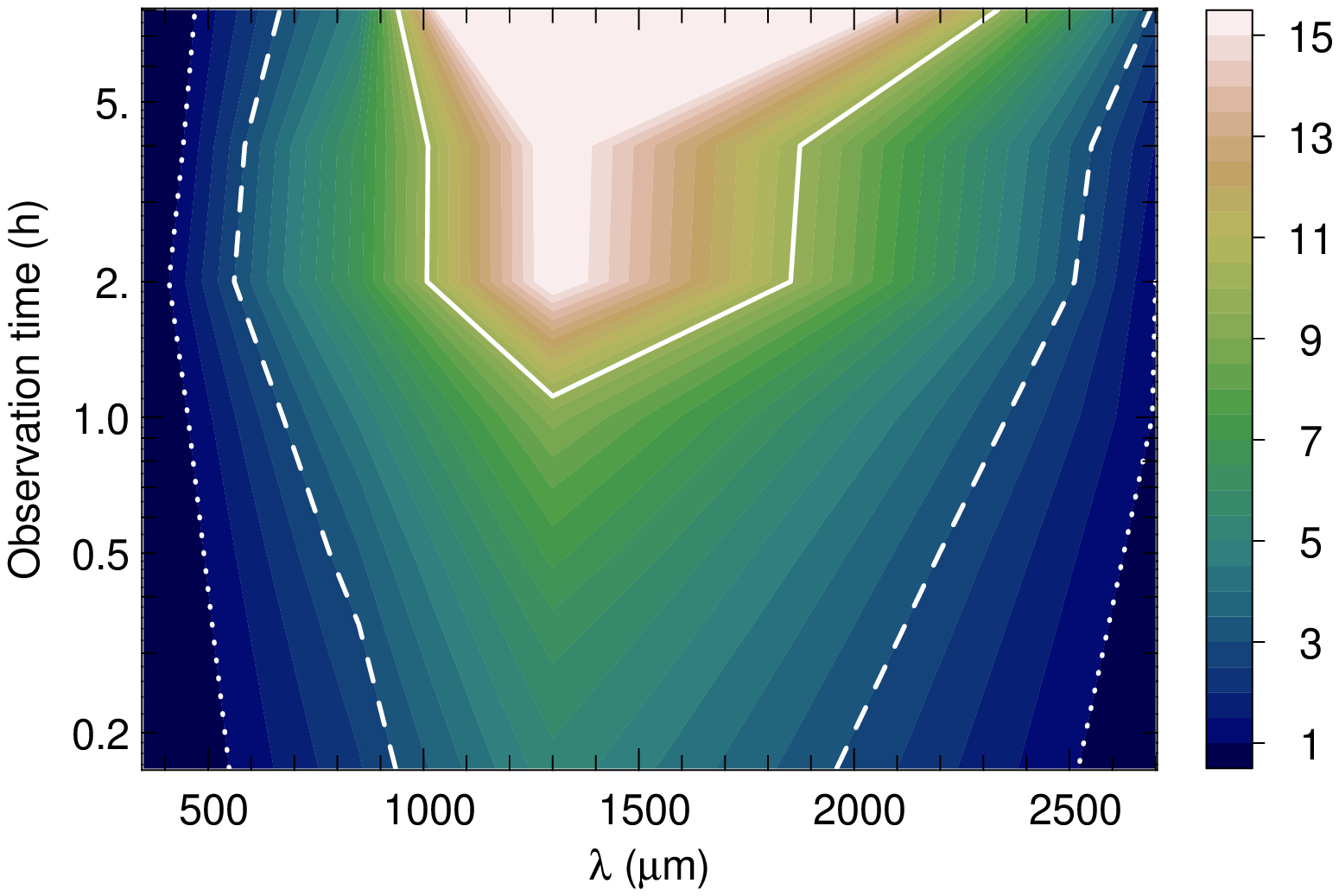}
}
\caption{Same as Fig.~\ref{Fig:Contrast} for the outer disk only, with phase noise included. Left: $M_\mathrm{p}=1\ M_\mathrm{J}$, right: $M_\mathrm{p}=5\ M_\mathrm{J}$.}
\label{Fig:Contrast_PhaseNoise}
\end{figure*}

The left columns of Figs.\,\ref{Fig:1Mj_PhaseNoise} and \ref{Fig:5Mj_PhaseNoise} show the effect of phase noise on the ALMA simulated images for both planet masses, to be compared to the center columns of Figs.\,\ref{Fig:1Mj_CASA} and \ref{Fig:5Mj_CASA}, computed without phase noise. Note that the absolute flux scale is different in both sets of figures. As expected, the image deterioration is strongest for the shortest wavelengths, to the point where the disk is no longer visible at 350\,$\mu$m. Despite some degradation at 850\,$\mu$m due to the higher noise level, both the disk and the gap are still well detected, in fact with a very similar disk-to-gap contrast to the images without phase noise. The image at 1.3\,mm is virtually unaffected by phase noise, and this wavelength appears to be the best to recover all the details of the disk. Figure\,\ref{Fig:Contrast_PhaseNoise} shows signal-to-noise maps in the outer disk in the images with phase noise for the 1 and $5\,M_\mathrm{J}$ planets, to be compared with the bottom panels of Fig.\,\ref{Fig:Contrast}. Phase noise naturally reduces the $S/N$ more strongly as wavelength decreases, whatever the integration time. The $S/N$ is almost unchanged at 2.7\,mm, where it was already very low, slightly reduced at 1.3\,mm with values $\sim$7--9 with an 1\,h integration time, degraded at 850\,$\mu$m to $\sim$3--4, and dramatically reduced to values well below unity at 350\,$\mu$m. Note that in the presence of phase noise, integrating longer does not improve the image, especially at short wavelengths. In particular, Fig.\,\ref{Fig:Contrast_PhaseNoise} shows that the $S/N$ at 350\,$\mu$m even decreases above 2\,h of integration.

While the detection of planet gaps is still possible at 850\,$\mu$m and good at 1.3\,mm, the differential signal-to-noise degradation hinders the possibility of using multi-wavelength observations to characterize the gap-opening planet. In particular, using the lowest wavelength to help separate disks with planet gaps from transition disks is no longer possible with the effect of phase noise. However, this can be mitigated by the real-time correction that will be available thanks to the use of WVRs (see Sect.\,\ref{sec:CASA}).

In practice, it is expected that observations at short wavelengths, and especially at 350\,$\mu$m, will only be performed in very dry weather. We now consider the best 25\% and 10\% of sky conditions and present the corresponding images, computed with respectively PWV = 0.58 and 0.30\,mm, in the center and right columns of Figs.\,\ref{Fig:1Mj_PhaseNoise} and \ref{Fig:5Mj_PhaseNoise}. As mentioned earlier, the effect of phase noise is moderate-to-negligible at 850\,$\mu$m and 1.3\,mm, respectively, and drier conditions mostly increase the $S/N$ thanks to the associated decrease of the thermal noise. At 350\,$\mu$m, on the other hand, even though decreasing the PWV to 0.58\,mm barely brings the disk out the noise, the improvement is dramatic in the driest case considered here, with PWV = 0.30\,mm, where both the disk and the gap are clearly detected with a contrast similar to the 850\,$\mu$m image. In the best 10\% of sky conditions, both the thermal and phase noises are so low that the shortest wavelengths are accessible without any necessity for correction and can provide the data needed to discriminate transition disks from disks with planet gaps even if the WVRs were not efficient. In such weather, the use of fully-functional WVRs and fast-switching will still improve the $S/N$.

\subsection{Varying disk inclination}
\label{sec:Inclination}

\begin{figure}
\centering
\resizebox{\hsize}{!}{
\includegraphics{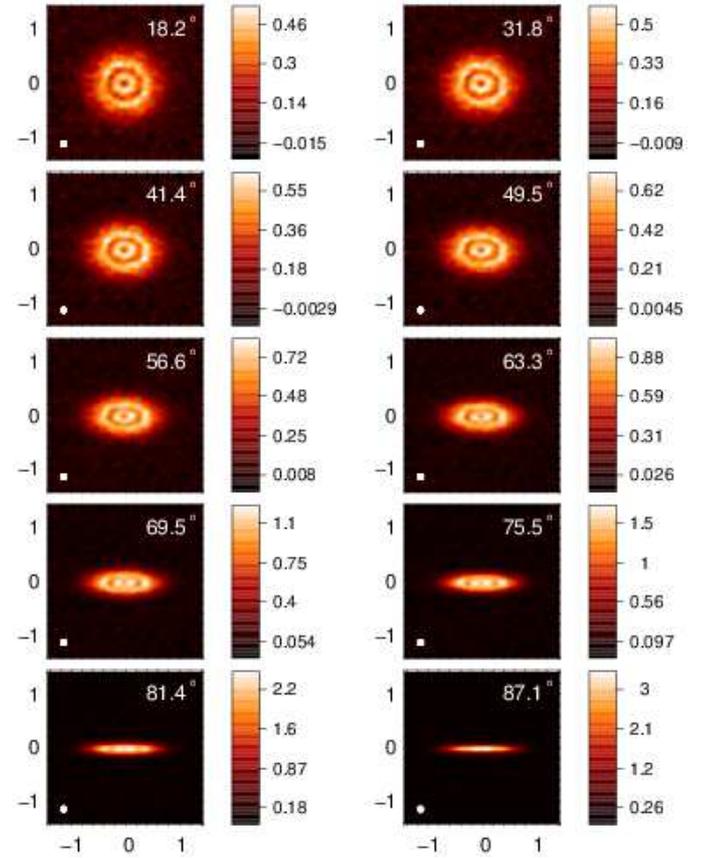}
}
\caption{Simulated observations of disks of varying inclination for the 1\,$M_\mathrm{J}$ planet in the dynamic case at $\lambda=850\ \mu$m with an integration time of 1\,h. The scale on each image is in arcseconds, with the beam size represented at its bottom left corner, the disk inclination in the upper right corner, and the colorbar gives the flux in mJy/beam.}
\label{Fig:Inclination}
\end{figure}

Since disks are not always viewed pole-on, we now investigate the effect of different disk inclinations. The images shown in Fig.\,\ref{Fig:Inclination} are calculated at 10 inclinations equally spaced in cosine, i.e. for $\cos(i)=0.05$, 0.15, \ldots, 0.85 and 0.95. This corresponds to disks randomly oriented in three dimensions, with inclinations to the line-of-sight $i=18.2$, 31.8, 41.4, 49.5, 56.6, 63.3, 69.5, 75.5, 81.4 and $87.1^\circ$. The planet gap is very easily seen for all inclinations up to $i=75.5^\circ$, only hinted at for $i=81.4^\circ$, and not visible for $i=87.1^\circ$. This is because the dust component of the disk, responsible for the emission in the continuum, has efficiently settled to the midplane and is almost flat (see Fig.\,\ref{Fig:hydro}). There is thus no flared regions in the outer disk that would mask the gap at high inclinations. Only when the disk is seen nearly edge-on does the outer disk hide the inner regions. Gap detection is therefore robust with respect to disk inclination.

\subsection{Varying source distance and declination}
\label{sec:OtherSFRs}

\begin{figure}
\centering
\resizebox{.85\hsize}{!}{
\includegraphics{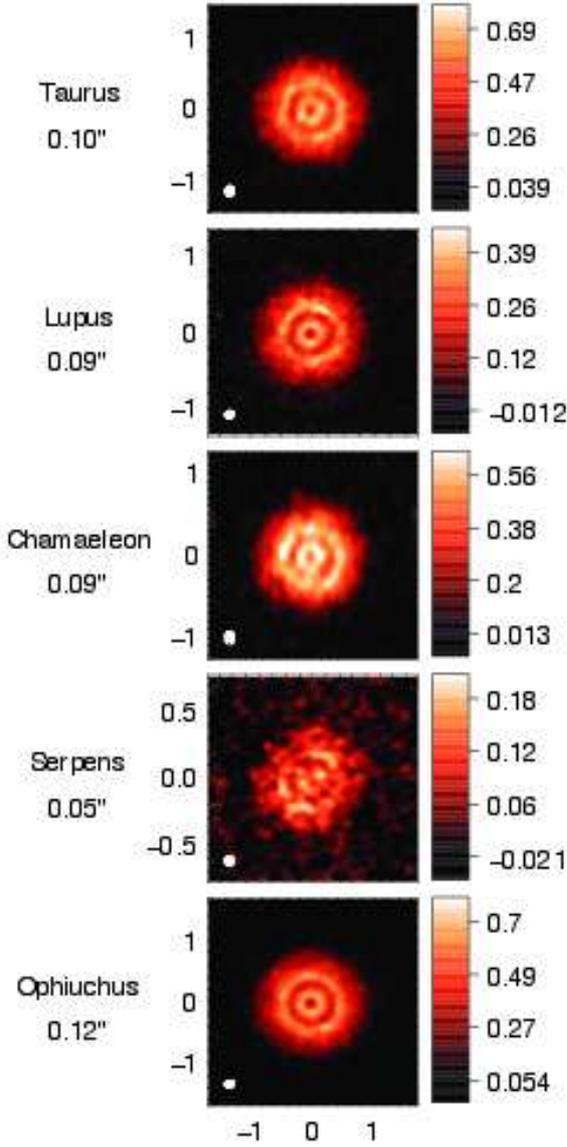}
}
\caption{Simulated observations of disks located in the star-forming regions listed in Table~\ref{Tab:SFR} for the 1\,$M_\mathrm{J}$ planet in the dynamic case at $\lambda=850\ \mu$m with an integration time of 1\,h. The angular resolution for each image is given on the left (it corresponds to an identical spatial resolution for all star forming regions), together with the region's name. The scale on each image is in arcseconds, with the beam size represented at its bottom left corner, and the colorbar gives the flux in mJy/beam.}
\label{Fig:SFR}
\end{figure}

ALMA simulated images are almost always shown in the most favorable case of a source passing through the zenith at ALMA's latitude (as is the case for our reference disk) and for a single distance. We present in Fig.\,\ref{Fig:SFR} simulated images where we vary the declination and distance of the source, taking values for specific star-forming regions listed in Table\,\ref{Tab:SFR}. The images are shown for the $1\,M_\mathrm{J}$ planet at 850\,$\mu$m and for a 1\,h observing time, without phase noise. In order to compare the gap detectability with our reference disk at 140\,pc observed with an angular resolution of $0.10''$, we chose to keep the same corresponding linear resolution of 14\,AU at the distance to each star-forming region and set the angular resolution accordingly.

\begin{table}
\caption{Declinations and distances of the specific star-forming regions used in our ALMA simulated images.}
\label{Tab:SFR}
\begin{center}
\begin{tabular}{lll}\hline\hline
star-forming region & $\delta$ ($^\circ$) & $d$ (pc) \\ \hline
Reference & $-$23 & 140 \\[.5em]
Taurus & +25 & 140 \\
Serpens & +01 & 260 \\
Ophiuchus & $-$24 & 120 \\
Lupus (I) & $-$34 & 150 \\
Chamaeleon (I) & $-$77 & 160 \\ \hline
\end{tabular}
\end{center}
\end{table}

For the most distant disks in Serpens at 260\,pc, the small angular resolution required ($0.05''$) is obtained by an extended array configuration, resulting in a low signal-to-noise ratio (see Sect.\,\ref{sec:ALMApredictions}). The gap is nonetheless detected, but characterizing the planet responsible for it will likely be impractical with these observing parameters. The other four star-forming regions are approximately twice as close, with similar distances of 120 to 160\,pc, corresponding to required angular resolutions of 0.09--$0.12''$. These are close to our optimal value, and naturally provide images where the gap is clearly seen.

Surprisingly, the source's declination has very little influence on the gap detectability, even for disks culminating at a rather low elevation, like in Taurus ($42^\circ$) or Chamaeleon ($36^\circ$). The beam is the most elongated for the latter (and as a result samples a larger area of the source, making the disk appear brighter) but not enough to smear out the sharp gap edges. This very good performance compared to disks reaching high elevations is due to the excellent $uv$ coverage that ALMA provides even in an observing time as short as 1\,h and ensures that the disk-to-gap contrast stays high for a large range of source declinations.

The prospect of detecting planet gaps with ALMA in nearby star-forming regions therefore seems excellent. Ophiuchus is obviously most favorable as it is the closest star-forming region and it culminates very close to the zenith at the location of ALMA. 

\subsection{Pushing ALMA further}
\label{sec:ALMAlimits}

\begin{figure}
\centering
\resizebox{.75\hsize}{!}{
\includegraphics{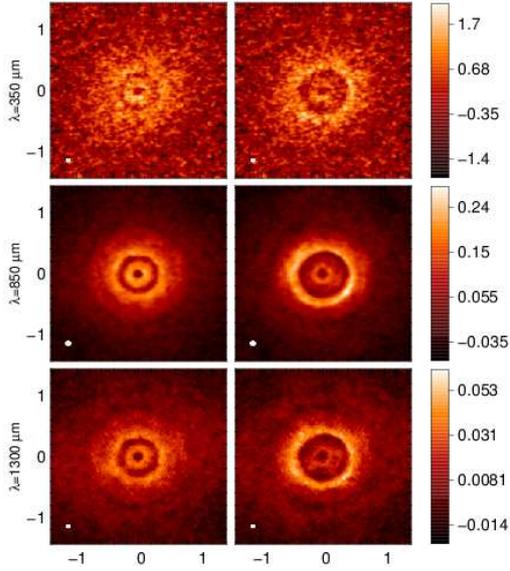}
}
\caption{Simulated observation for our reference disk with an integration time of 8\,h and an angular resolution of $0.05''$. Left: $M_\mathrm{p}=1\ M_\mathrm{J}$, right: $M_\mathrm{p}=5\ M_\mathrm{J}$. From top to bottom, $\lambda=350\,\mu$m, 850\,$\mu$m and 1.3\,mm. The scale on each image is in arcseconds, with the beam size represented at its bottom left corner, and the colorbar gives the flux in mJy/beam.}
\label{Fig:ALMAlimits}
\end{figure}

The detection of planet signatures is a key science driver for ALMA, and one can reasonably expect that it will be possible to commit a substantial amount of time on promising sources, which will be necessary as one wants to reach high angular resolution and high $S/N$. In the context of trying to obtain the very best of what ALMA can offer, we computed simulated ALMA images of our reference disk for both planet masses with a $0.05''$ angular resolution, an integration time of 8\,h, and no phase noise (assuming the WVRs will be able to correct most of it), shown in Fig.\,\ref{Fig:ALMAlimits}. As expected, the images for both planet masses show much more sharply defined gaps, allowing a more accurate measure of their width, than for the observing modes discussed above. For the 1\,$M_\mathrm{J}$ planet, the improvement in the disk-to-gap contrast is particularly striking at 350\,$\mu$m, increasing from $\sim$1.4 to $\sim$1.8, but no additional features are recovered. On the other hand, new details now appear for the 5\,$M_\mathrm{J}$ planet. At 850\,$\mu$m and 1.3\,mm, emission from the corotating grains can be seen but unfortunately remains too faint to be used as a constraint on the planet mass. However, the brightest parts of the outer disk now reveal the spiral wave, especially at 850\,$\mu$m, which can help constrain the planet's angular position.

Spiral features have already been detected in visible or near-infrared scattered-light images of disks a few hundred AU from their star, e.g. \object{AB~Aur} \citep{Grady1999,Fukagawa2004}, \object{HD\,100546} \citep{Grady2001} and \object{HD\,141569A} \citep{Clampin2003}. Subsequent mm observations of the well-studied \object{AB~Aur} disk \citep{Pietu2005,Lin2006} estimated the Toomre $Q$ parameter well above unity, ruling out the gravitational instability scenario that was proposed previously and favoring excitation by a giant planet at several tens of AU to $\sim$100~AU. More recently, \citet{Hashimoto2011} obtained high-resolution $H$-band polarized intensity images of the \object{AB~Aur} inner disk and detected two rings separated by a gap at 80~AU, possibly caused by a Jupiter-mass planet, and concluded that the disk is probably in an early and active phase of planet formation. The first example of a direct constraint from observed spiral features comes from \citet{Muto2012} who detected two spirals in $H$-band polarized intensity images of the \object{HD\,135344B} disk: using spiral density wave theory, they infer the presence of two unseen planets of $\sim$0.5\,$M_\mathrm{J}$. ALMA, observing at longer wavelengths, will probe structures closer to the midplane and bring additional constraints to such studies.

\begin{figure}
\centering
\resizebox{.75\hsize}{!}{
\includegraphics{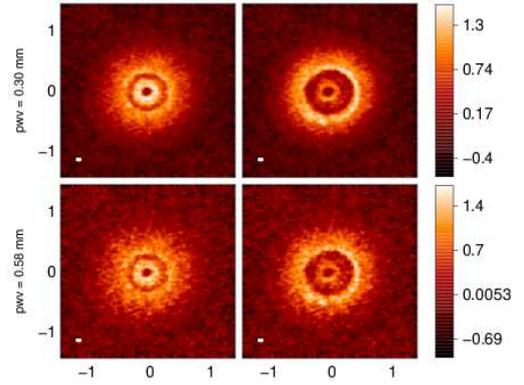}
}
\caption{Same as Fig.\,\ref{Fig:ALMAlimits} for $\lambda=350\,\mu$m only. Left: $M_\mathrm{p}=1\ M_\mathrm{J}$, right: $M_\mathrm{p}=5\ M_\mathrm{J}$. Top: PWV\,=\,0.30\,mm, bottom: PWV\,=\,0.58\,mm.}
\label{Fig:ALMAlimitsPWV}
\end{figure}

\begin{figure*}
\resizebox{.5\hsize}{!}{
\includegraphics{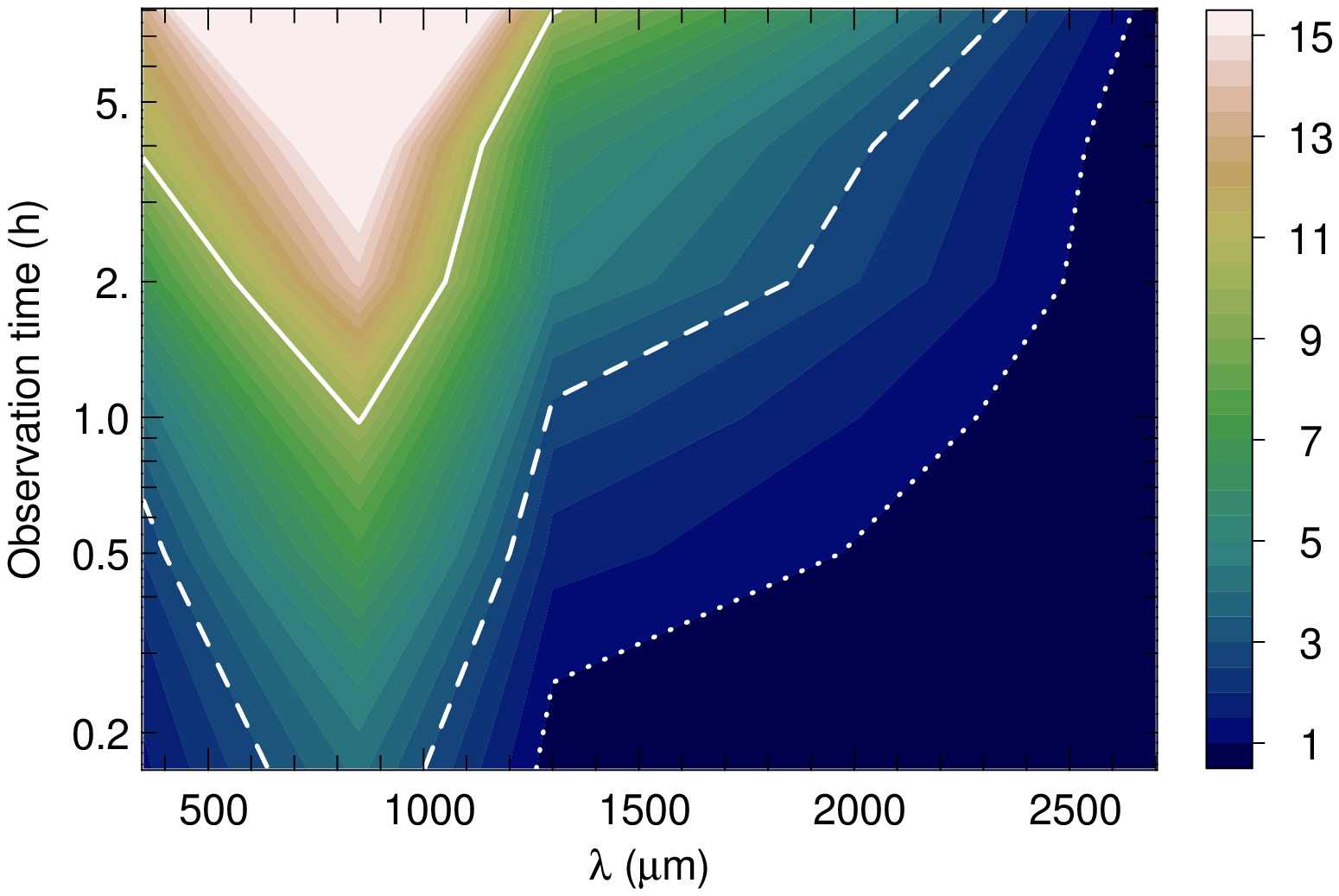}
}
\resizebox{.5\hsize}{!}{
\includegraphics{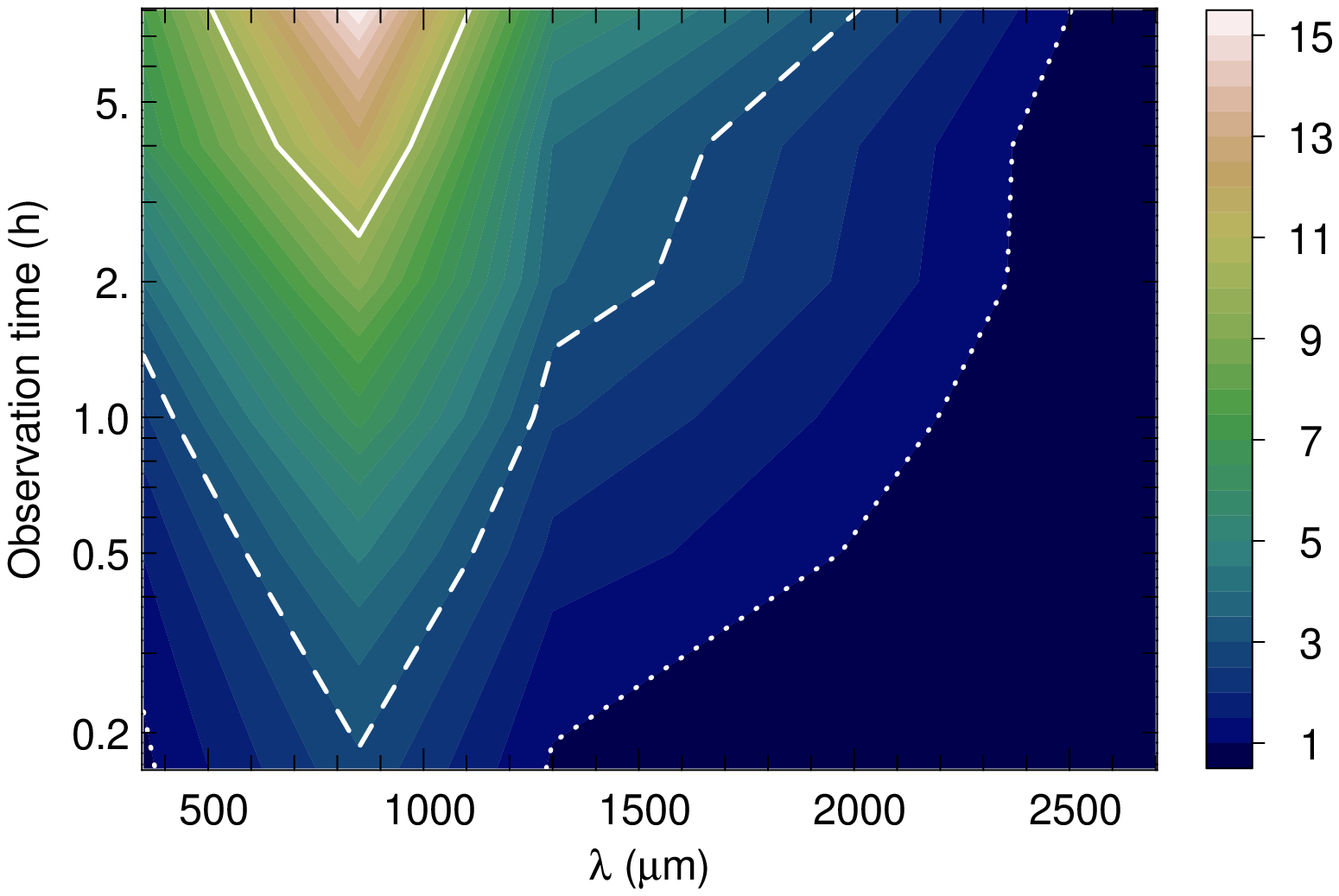}
}
\caption{Same as Fig.~\ref{Fig:Contrast} for the inner disk only, with an angular resolution of $0.05''$ and a PWV of 0.30\,mm. Left: $M_\mathrm{p}=1\ M_\mathrm{J}$, right: $M_\mathrm{p}=5\ M_\mathrm{J}$.}
\label{Fig:Contrast_PWV0.3}
\end{figure*}

To push ALMA performances even further, we also computed simulated images assuming drier conditions giving the 10 and 25\% best values of the sky opacity, corresponding to a PWV of 0.30 and 0.58\,mm, respectively (see Table\,\ref{Tab:SkyConditions}). These are shown in Fig.\,\ref{Fig:ALMAlimitsPWV} for $\lambda=350\,\mu$m only (images at longer wavelengths are virtually unaffected). For these lower opacities, the thermal noise (as well as the phase noise, not included here) is strongly reduced, resulting in images with a much improved $S/N$ at the same angular resolution. Figure\,\ref{Fig:Contrast_PWV0.3} shows signal-to-noise maps in the inner disk in the images with an angular resolution of $0.05''$ and a PWV of 0.30\,mm for the 1 and $5\,M_\mathrm{J}$ planets. $S/N$ values are lower than those in the top panels of Fig.\,\ref{Fig:Contrast} in spite of better sky conditions because the smaller beam collects less signal in the same amount of time. $S/N$ values above 3 are nevertheless easily reached in one hour, and Fig.\,\ref{Fig:Contrast_PWV0.3} can be used to determine the integration time needed to reach a given $S/N$ at a given wavelength.

Pushing ALMA towards its limits therefore seems worthwhile to better characterize sources of interest. Here, 850\,$\mu$m appears to be the most favorable wavelength in average sky conditions, recovering the most details in the disk for both planet masses. In the driest weather, observations at the shortest wavelength of 350\,$\mu$m are optimal and allow the detection of wavelength-dependent features such as the brightness of the inner disk. However, Fig.\,\ref{Fig:Contrast_PWV0.3} shows that 850\,$\mu$m remains the wavelength giving the best $S/N$ for our disks, even in the best 10\% of sky conditions.

\section{Conclusion}
\label{sec:Conclusion}

In a previous study \citepalias[see][]{Fouchet10}, we have run 3D hydrodynamic simulations of the gaps carved in a 0.02\,$M_\odot$ disk of gas and dust by 1 and 5\,$M_\mathrm{J}$ planets at 40\,AU from the 1\,$M_\odot$ central star. We followed consistently the dynamics of dust grains in the range of sizes contributing the most to the ALMA wavelengths and described the distinctive, sharper features in the dust phase. In this paper, we now provide predictions of observations with ALMA and examine the detectability of these planetary gaps. We first produce raw synthetic images by applying the radiative transfer code \mcfost\ to the hydrodynamic results. We produce two kinds of images: the ``dynamic'' case where we use the true spatial distribution of dust of different grain sizes obtained from the hydrodynamic simulations and the ``well-mixed'' case where we simply use the spatial distribution of gas and assume that the dust follows it in order to check the validity of the ``well-mixed'' hypothesis.

We then produce simulated ALMA images using the CASA software package. We investigate the choice of integration time, wavelength and angular resolution. We show that an integration time of 1\,h is enough to ensure a firm detection of our planet gaps and that the gain in signal-to-noise as we increase the integration time does not warrant the necessary extra observing time on a facility that will be heavily oversubscribed. We also show that an angular resolution of $0.10''$ is optimal to resolve the gap with the best contrast in most nearby star-forming regions. We then show that the use of small or large wavelengths combined with these optimal observing parameters give poor to bad results. In average sky conditions, we find the wavelength giving the best signal-to-noise ratio to be 850\,$\mu$m when phase noise is neglected or 1.3\,mm, with slightly lower performance, when it is included. However, the use of WVRs should provide real-time correction of phase noise and favor 850\,$\mu$m as the wavelength of choice. In the driest weather, phase noise is negligible.

A number of previous studies of planet gaps \citep[e.g.][]{W02,W05}, and even recent work on Rossby vortices in protoplanetary disks \citep{Regaly2012}, have relied on hydrodynamic simulations of 2D gas-only disks to produce simulations of ALMA observations by reconstructing a three-dimensional distribution of dust that follows that of the gas in the midplane and is in hydrostatic equilibrium in the vertical direction before feeding it to a radiative transfer code. In addition to neglecting the vertical dust settling, this approach also results in smoother dust features, leading to more pessimistic predictions for the detectability of the disk features. The comparison of our simulated images in the ``well-mixed'' and ``dynamic'' cases show that they are very different and that the more realistic case of including a self-consistent treatment of the dust dynamics produces much sharper features and more clearly defined gaps. We would like to stress here that investigators need to be cautious when assessing the observability of a particular structure in their source from approximate methods and that a procedure similar to our ``dynamic'' case should be used whenever possible.

With our ``dynamic'' approach, we find that the gap carved by a 1\,$M_\mathrm{J}$ planet orbiting 40\,AU from its star is easily detected. Images for the 5\,$M_\mathrm{J}$ planet show a high contrast between the outer disk and the deeper and wider gap, but the fainter inner disk is barely seen with our standard observing parameters. In order to detect the regions interior to the gap and remove the possible confusion with transition disks with large inner holes, one needs to observe at shorter wavelengths in the driest conditions possible or use a combination of longer integration time and higher angular resolution. In any case, multi-wavelength followup observations of detected gaps is advisable for a characterization of the planet.

This work demonstrates that the gap carved by a moderately massive planet of one Jupiter mass at large orbital radii will be well within reach of ALMA. While the image quality naturally decreases with increasing distance, we find that the ability to observe planet gaps is little affected by the source declination or inclination to the line-of-sight. Although few planets on wide orbits have been detected so far due to the biases of current techniques, their existence is supported by the more recent planet formation models as well as by observational constraints on planet populations. ALMA, probing a new part of the orbital parameter space, has the potential to unveil more of them in nearby star-forming regions.

\begin{acknowledgements}
We thank Fr\'ed\'eric Gueth for enlightening discussions about simulating the effects of phase noise in ALMA reconstructed images and Remy Indebetouw, Scott Schnee and Juergen Ott for help on the use of the CASA ALMA simulator. This research was partially supported by the Programme National de Physique Stellaire, the Programme National de Plan\'etologie and the Action Sp\'ecifique ALMA of CNRS/INSU, France, and the Agence Nationale de la Recherche (ANR) of France through contracts ANR-07-BLAN-0221 and ANR-2010-JCJC-0504-01. We acknowledge funding from the European Commission's 7$^\mathrm{th}$ Framework Program (contract PERG06-GA-2009-256513) and from the Millenium Science Initiative (ICM) of the Chilean ministry of Economy (Nucleus P10-022-F, ``Disks with ALMA''). Simulations presented in this work were run at the Service Commun de Calcul Intensif (SCCI) de l'Observatoire de Grenoble, France. Images in Fig.\,\ref{Fig:hydro} were made with SPLASH \citep{Price2007}, all other figures were made with the freeware {\sf Yorick}.

\end{acknowledgements}

\bibliographystyle{aa}
\bibliography{aa-2012-18806}

\Online

\begin{appendix}
\section{Dust density sampling in the MCFOST grid}
\label{App:N_SPH}

\begin{figure}
\centering
\resizebox{\hsize}{!}{
\includegraphics{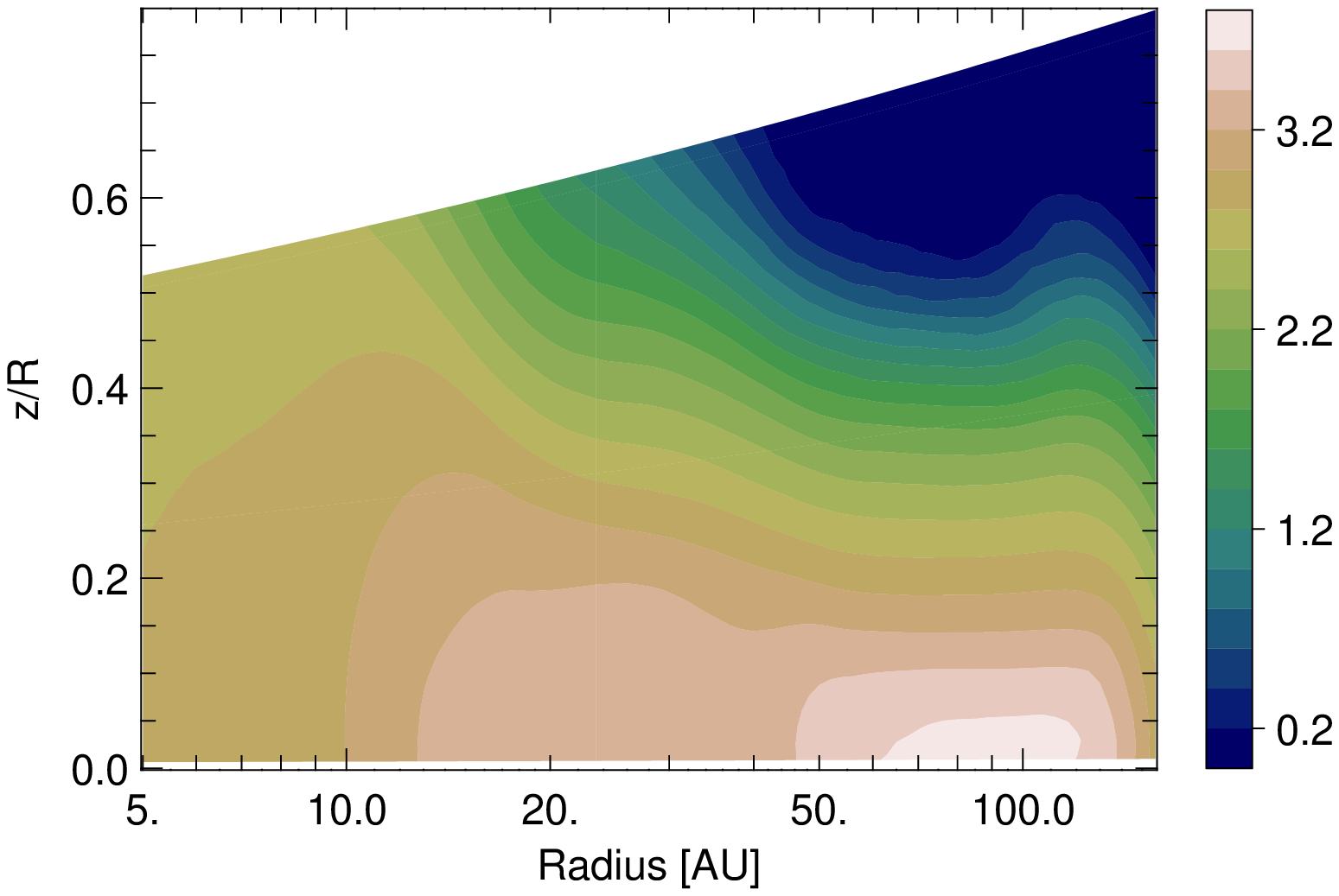}
}
\caption{Map of the logarithm of the number of SPH particles per \mcfost\ grid cell in the meridian $(R,z/R)$ plane for the 1\,$M_\mathrm{J}$ planet.}
\label{Fig:N_SPHrz}
\end{figure}

\begin{figure}
\centering
\resizebox{\hsize}{!}{
\includegraphics{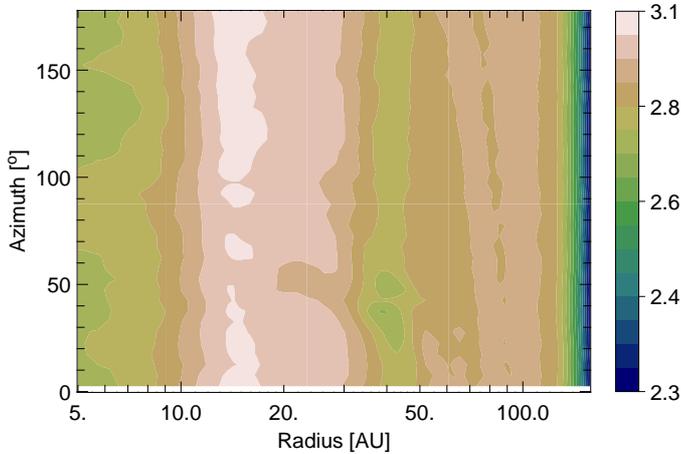}
}
\caption{Same as Fig.~\ref{Fig:N_SPHrz} in the horizontal $(R,\phi)$ plane.}
\label{Fig:N_SPHrtheta}
\end{figure}

In order for \mcfost\ to produce accurate synthetic images from the output of SPH simulations, the dust density distribution must be sufficiently sampled with SPH particles in each cell of the radiative transfer grid. Figures~\ref{Fig:N_SPHrz} and \ref{Fig:N_SPHrtheta} show 2D maps of the logarithm of the number of SPH particles per grid cell in the $(R,z/R)$ and $(R,\phi)$ planes respectively. In most of the disk, the number of SPH particles contributing to the density in any grid cell varies from a few hundred to a few thousand (as a reference, SPH codes typically use a few tens of neighbouring particles to compute physical quantities). Only in the upper layers at large distances from the star does this number fall to values of a few, because there is very little material in these regions. Indeed, as can be seen in Fig.~\ref{Fig:hydro}, dust grains 100\,$\mu$m in size and larger settle efficiently to the midplane and smaller grains follow the distribution of the gas phase, which has a curved outer-rim and therefore a very low density in these regions. Both maps show smooth variations in the radial and vertical directions, whereas azimuthal variations are negligible (note the different colorscale in Figs.~\ref{Fig:N_SPHrz} and \ref{Fig:N_SPHrtheta}), and demonstrate the adequate sampling of the dust distribution for the radiative transfer calculations.

\section{Impact of the disk inner radius on the temperature structure}
\label{App:Tstruct}

\begin{figure}
\includegraphics[width=\hsize]{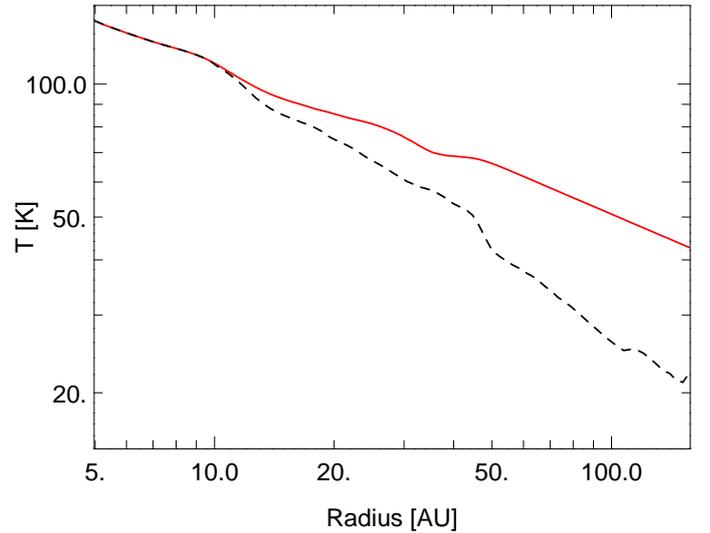}
\caption{Radial temperature profile at the disk surface (solid red line) and in the disk midplane (black dashed line).}
\label{fig:verif_T}
\end{figure}

The inner radius of the disk models is set to 4\,AU in the SPH calculations, much larger than the characteristic inner radius of most T~Tauri disks (usually located at the dust sublimation radius $\approx$ 0.05\,AU). The resulting temperature structure is therefore incorrect in the central parts of the model, where the dust is directly heated by the star. Our disk models remain optically thick in the radial direction however. As a consequence the temperature in the outer parts is computed correctly in these regions, as the heating of the disk midplane (where most of the millimeter emission is coming from) is due to reprocessing of the stellar light absorbed and scattered by the disk surface. In these regions, the transfer of radiation is mostly vertical and the temperature structure does not depend on the details of the inner disk. Figure~\ref{fig:verif_T} shows the radial profile of the surface and midplane temperatures. In the inner region, the dust is heated directly by the stellar radiation and both temperatures are equal. They start to decouple at $r\approx10$\,AU, when the disk midplane no longer sees the star directly. The temperature structure and corresponding millimeter images can therefore be considered accurate from $r>15$\,AU, corresponding to 0.1'' in most of our simulations.

\section{Additional figures}
\label{App:Figs}

\begin{figure*}
\centering
\resizebox{.85\hsize}{!}{
\includegraphics[angle=-90]{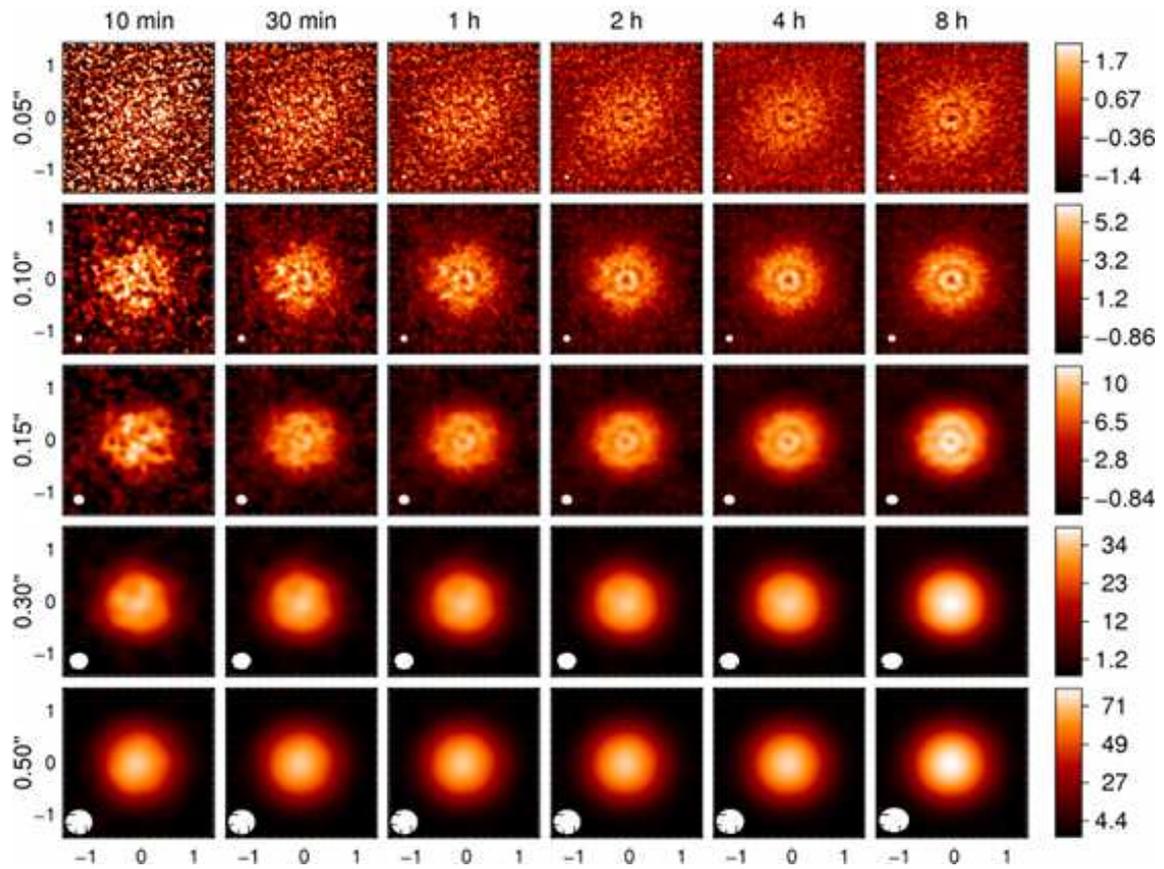}
}
\caption{Same as Fig.\,\ref{Fig:IntTimeResol} at $\lambda=350\ \mu$m.}
\label{Fig:IntTimeResol350}
\end{figure*}

\begin{figure*}
\centering
\resizebox{.85\hsize}{!}{
\includegraphics[angle=-90]{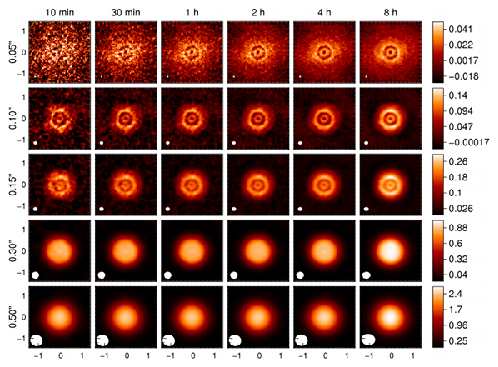}
}
\caption{Same as Fig.\,\ref{Fig:IntTimeResol} at $\lambda=1.3$~mm.}
\label{Fig:IntTimeResol1300}
\end{figure*}

\begin{figure*}
\centering
\resizebox{.85\hsize}{!}{
\includegraphics[angle=-90]{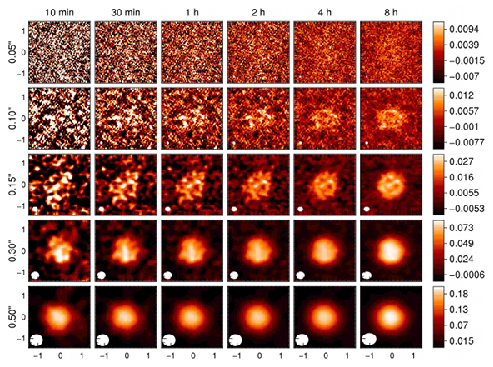}
}
\caption{Same as Fig.\,\ref{Fig:IntTimeResol} at $\lambda=2.7$~mm.}
\label{Fig:IntTimeResol2700}
\end{figure*}

We present here figures similar to Fig.\,\ref{Fig:IntTimeResol} for other wavelengths: $\lambda=350\ \mu$m in Fig.\,\ref{Fig:IntTimeResol350}, $\lambda=1.3$~mm in Fig.\,\ref{Fig:IntTimeResol1300}, and $\lambda=2.7$~mm in Fig.\,\ref{Fig:IntTimeResol2700}.

\end{appendix}

\end{document}